\documentclass[acmsmall]{acmart}

\acmJournal{TOSEM}

\usepackage{adjustbox}
\usepackage{algorithm}
\usepackage{alphalph}
\usepackage{amsfonts}
\usepackage{array}
\usepackage{booktabs}
\usepackage{diagbox}
\usepackage{enumerate}
\usepackage{enumitem}
\usepackage{fontawesome5}
\usepackage{graphicx}
\usepackage{listings}
\usepackage{float}
\usepackage{longtable}
\usepackage{makecell}
\usepackage{multirow}
\usepackage{nicematrix}
\usepackage{pdflscape}
\usepackage{rotating}
\usepackage{tabularx}
\usepackage{tcolorbox}
\usepackage{xcolor}
\usepackage{xspace}
\usepackage{xurl}
\usepackage[none]{hyphenat}
\usepackage{subcaption}
\tcbuselibrary{skins, breakable}

\makeatletter
\newcommand{\ie}{\emph{i.e.}\@ifnextchar.{\@gobble}{\@\xspace}}
\newcommand{\eg}{\emph{e.g.}\@ifnextchar.{\@gobble}{\@\xspace}}
\newcommand{\vs}{\emph{vs.}\@ifnextchar.{\@gobble}{\@\xspace}}
\newcommand{\etc}{\emph{etc}\@ifnextchar.{}{.\@\xspace}}
\newcommand{\aka}{\emph{a.k.a}\@ifnextchar.{}{.\@\xspace}}
\makeatother

\newcommand{\collectdate}{Jan 6th, 2026}

\newcolumntype{C}[1]{>{\centering\arraybackslash}p{#1}}

\newtcolorbox{mybox}[2][]{
top=0.15in,left=4pt,right=4pt,bottom=4pt,
fonttitle=\bfseries,
colbacktitle=gray,
colback=gray!5,
colframe=gray!40!black,
enhanced,
attach boxed title to top left={xshift=0em,yshift=-\tcboxedtitleheight/2},
boxed title style={size=small},
drop shadow={black!50!white},
title=#2,#1}

\newtcolorbox{finding}[1]{
  colframe=green!40!black,
  colback=green!5!white,
  coltitle=white,
  fonttitle=\bfseries,
  title={\faLightbulb\hspace{0.5em}\textbf{#1}},
  sharp corners,
  boxrule=1pt,
  boxsep=1mm,
  left=2mm,
  right=2mm,
  top=1mm,
  bottom=1mm,
  breakable
}

\begin{document}

\title{Towards Evaluation Engineering: An Empirical Study of ML Evaluation Harnesses in the Wild}

\author{Zhimin Zhao}
\affiliation{\institution{Software Analysis and Intelligence Lab (SAIL), School of Computing, Queen's University}\city{Kingston}\state{ON}\country{Canada}}
\email{z.zhao@queensu.ca}

\author{Zehao Wang}
\affiliation{\institution{Concordia University}\city{Montreal}\state{QC}\country{Canada}}
\email{w\_zeha@encs.concordia.ca}

\author{Abdul Ali Bangash}
\affiliation{\institution{Lahore University of Management Sciences (LUMS)}\city{Lahore}\state{Punjab}\country{Pakistan}}
\email{bangash@ualberta.ca}

\author{Bram Adams}
\affiliation{\institution{Software Analysis and Intelligence Lab (SAIL), School of Computing, Queen's University}\city{Kingston}\state{ON}\country{Canada}}
\email{bram.adams@queensu.ca}

\author{Ahmed E. Hassan}
\affiliation{\institution{Software Analysis and Intelligence Lab (SAIL), School of Computing, Queen's University}\city{Kingston}\state{ON}\country{Canada}}
\email{ahmed@cs.queensu.ca}

\begin{abstract}
Evaluation harnesses are software systems that orchestrate model evaluation by managing model invocation, data loading, metric computation, and result reporting. Despite their critical role in machine learning infrastructure, their operational challenges and engineering concerns have received limited attention so far. We present an empirical study of $57$ evaluation harnesses, deriving a five-stage harness model and classifying $16{,}560$ issues by workflow stage and root cause. Most harness operational challenges concentrate in the Specification stage ($41.4\%$ of issues), where harnesses integrate external models, datasets, and scoring judges. The three most frequent root causes of operational challenges are unimplemented features ($24.3\%$), documentation gaps ($20.3\%$), and missing input validation ($17.2\%$), which together account for $61.7\%$ of classified issues, spanning both defects in existing functionality and capability gaps that block intended workflows. Root causes also vary by workflow stage: environment incompatibility and external dependency breakage account for $36.2\%$ of provisioning issues, whereas algorithmic error ($25.9\%$) and validation gap ($22.5\%$) dominate assessment issues. Together, these contributions establish an empirical foundation for treating evaluation engineering as a distinct software engineering concern.
\end{abstract}



\maketitle

\section{Introduction}
\label{sec:introduction}

\begin{figure}[!t]
\centering
\includegraphics[width=\columnwidth]{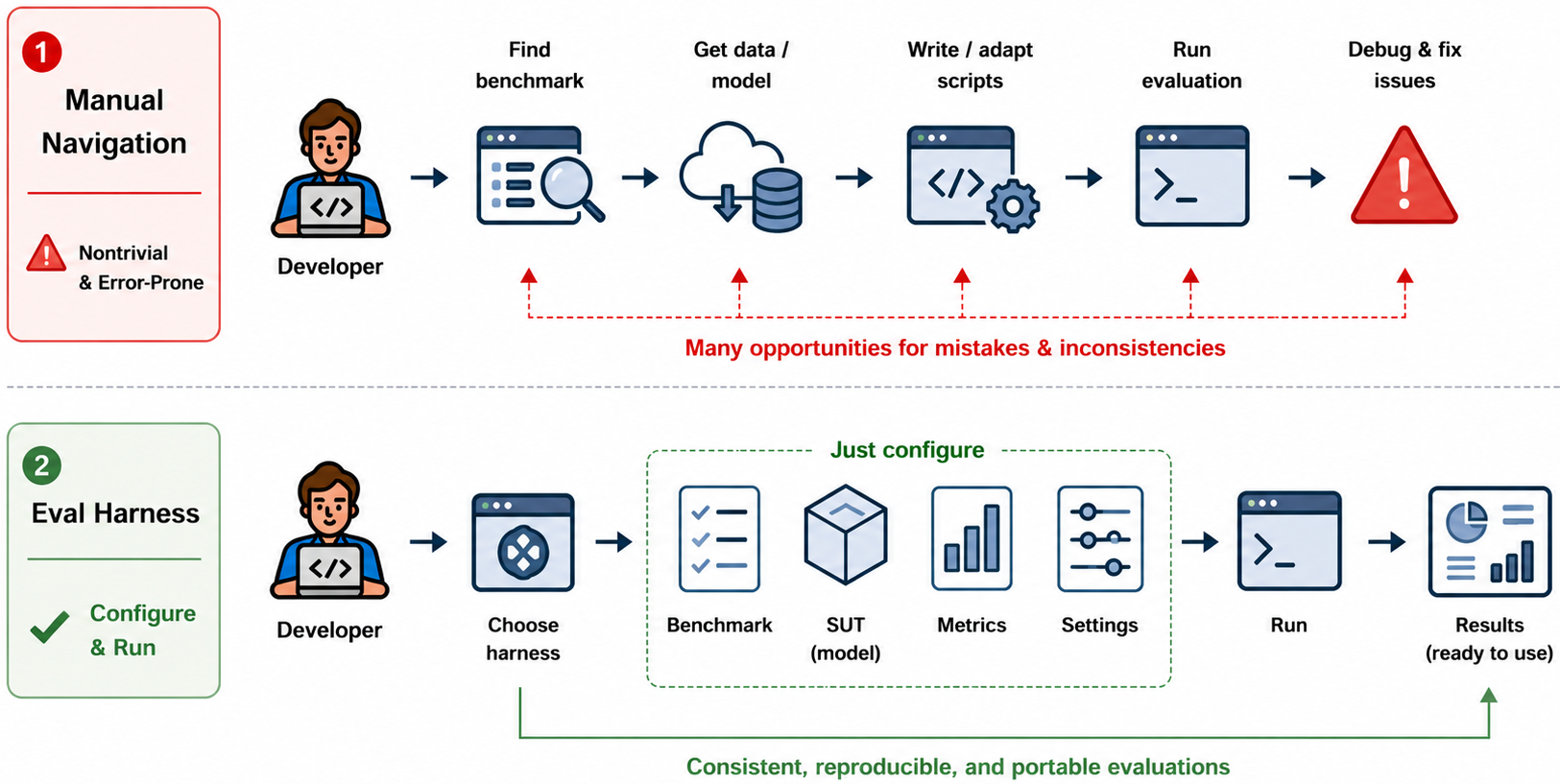}
\caption{From manual benchmark evaluation to configuration-driven evaluation harness workflow.}
\label{fig:intro_comparison}
\end{figure}

Machine learning (ML) model evaluation underpins progress in artificial intelligence (AI) research and development. Reliable evaluation depends not only on well-designed metrics and benchmarks, but also on the software infrastructure that executes them. To manage this infrastructure, the ML community has built \emph{evaluation harnesses}, \ie, systems that orchestrate model invocation, data loading, metric computation, and result reporting across diverse evaluation scenarios. Examples include LM Eval~\cite{eval-harness} and HELM~\cite{liang2022holistic}. As Figure~\ref{fig:intro_comparison} illustrates, harnesses replace ad hoc benchmark evaluation with configuration-driven evaluation workflows.

Despite this central role, however, no prior software engineering (SE) work has studied evaluation harnesses as software products, examining their operational workflows, the root causes of user challenges, and the engineering decisions that shape harness reliability. Existing work examines ML evaluation from methodological perspectives, focusing on what metrics to compute~\cite{chang2024survey,zhou2024survey}, what capabilities to test~\cite{mondorf2024beyond,gallegos2024bias,cecchini2024holistic}, and what challenges arise in benchmark design~\cite{sainz2023nlp,singh2024evaluation,biderman2024lessons} (\S\ref{sec:background}). These studies address \emph{what} to evaluate but not \emph{what SE challenges arise when} evaluation is operationalized. We use the term \emph{evaluation engineering} (EvalEng) to refer to the SE concerns that arise in this operationalization, covering harness design, dependency management, scoring correctness, and result integrity.

To address this gap, we conduct an empirical study of evaluation harnesses as software products. We analyze documentation, perform local execution, and examine GitHub issue reports (bug reports, feature requests, and usage questions) from $57$ harnesses to extract a unified workflow model, identify where developers encounter friction, and categorize the root causes of the challenges they face. We investigate three research questions (RQs):
\begin{itemize}[leftmargin=*]
    \item \textbf{RQ1}: \emph{What is the operational workflow for evaluation harness execution across different ML domains?} We extract stages, steps, and concrete implementation strategies observed across harnesses, producing a hierarchical workflow model from environment setup through result reporting.
    \item \textbf{RQ2}: \emph{What are the root causes of operational challenges in evaluation harnesses?} We develop a root cause taxonomy from developer-reported GitHub issues, covering both software defects and capability gaps that block harness operation, and characterize the prevalence of each root cause across evaluation harnesses.
    \item \textbf{RQ3}: \emph{How do operational root cause distributions vary across evaluation workflow stages?} We map root causes onto the workflow model from RQ1, showing how each root cause concentrates in specific stages and how stages differ in their failure composition.
\end{itemize}

We employ a four-stage methodology combining qualitative workflow extraction via open card sorting~\cite{spencer2009card} with large-scale GitHub issue mining~\cite{bhatia2023empirical}. First, we identify $57$ evaluation harnesses through curated sources and keyword-based GitHub search. Second, we extract a workflow model through iterative open card sorting of harness documentation and local execution, with constant comparison until theoretical saturation. Third, we mine $19{,}638$ GitHub issues from these harnesses. Fourth, we use LLM-based classifiers, calibrated against human consensus labels ($\kappa > 0.87$), to map issues onto workflow stages and root cause categories at scale.

Our analysis yields the following findings. First, \textbf{integrating external dependencies is the largest source of operational challenges}. The Specification stage, where harnesses load models, datasets, and scoring judges, accounts for $41.4\%$ of all issues. Within this stage, integration with remote model APIs (authentication failures, endpoint changes, and rate limits) accounts for $48.5\%$ of model preparation issues, and loading and accessing offline benchmark data (change in data availability, format mismatches, and preprocessing failures) accounts for $76.4\%$ of input preparation issues. Second, \textbf{capability gaps and documentation gaps are the most frequent root causes}: unimplemented features ($24.3\%$), documentation gaps ($20.3\%$), and missing input validation ($17.2\%$) together account for $61.7\%$ of all classified issues, while scoring errors ($8.3\%$) are less frequent than integration and usability failures, indicating that the dominant engineering burden in evaluation harnesses lies in operationalization rather than metric computation. Root cause distributions vary by workflow stage: environment incompatibility and external dependency breakage account for $36.2\%$ of provisioning issues, whereas algorithmic error ($25.9\%$) and validation gap ($22.5\%$) dominate assessment. Third, \textbf{harnesses show uneven adoption of production-oriented capabilities}: only $22.8\%$ quantify uncertainty around scores, and $8.8\%$ provide regression alerting to detect score degradation between runs.

This study contributes: (1) an \textbf{operational workflow model} comprising $5$ stages, $9$ steps, and $34$ strategies for ML model evaluation; (2) an \textbf{empirical mapping of operational engineering challenges} from $19{,}638$ GitHub issues across $57$ harnesses; (3) a \textbf{root cause taxonomy} of ten challenge categories, spanning both software defects and capability gaps, across $16{,}560$ classified issues; (4) \textbf{identification of engineering adoption gaps} (\ie, capabilities that most harnesses have not yet implemented or fully documented) in production-oriented areas such as uncertainty quantification and regression alerting. Together, these contributions establish an empirical foundation for EvalEng as a distinct SE concern, showing implications for harness developers, users, and researchers that we discuss in Section~\ref{sec:implications}.

The remainder of this paper is organized as follows. Section~\ref{sec:background} reviews background and related work. Section~\ref{sec:methodology} describes our four-stage methodology. Sections~\ref{sec:rq1-results},~\ref{sec:rq2-results}, and~\ref{sec:rq3-results} present the results for RQ1, RQ2, and RQ3, respectively. Section~\ref{sec:implications} discusses implications for harness developers, users, and researchers. Section~\ref{sec:threats} addresses threats to validity, and Section~\ref{sec:conclusion} concludes the paper.

\section{Background and Related Work}
\label{sec:background}

\subsection{Evaluation as the Foundation of ML Progress}
\label{sec:bg_foundation}

ML evaluation measures model performance on standardized tasks, enabling researchers to compare methods and track improvements. Recent work argues that verification asymmetry, the observation that validating solutions is fundamentally easier than generating them, determines which ML capabilities become tractable~\cite{zhao2026whycode,keles2025verifiability,noroozi2024networks,wei2024verifier,goldwasser2021interactive}. This asymmetry explains why ML advances rapidly on tasks with reliable verification infrastructure: competitive programming succeeded because test suites provide instant correctness feedback, mathematical reasoning progressed through symbolic verification, and code generation improved via executable unit tests. The pattern reveals a dependency: ML advancement relies on evaluation infrastructure that can reliably measure progress.

Well-documented challenges can affect the reliability of ML evaluation in practice: benchmark contamination (overlap between training data and evaluation data) inflates performance estimates~\cite{sainz2023nlp,yang2023rethinking,xu2024benchmark,singh2024evaluation}, unreported implementation details prevent reproducibility of evaluation results~\cite{singh2024evaluation,semmelrock2025reproducibility}, incompatible frameworks fragment cross-study comparison of model performance~\cite{maslej2024artificialintelligenceindexreport,biderman2024lessons}, annotation errors (incorrect human-provided labels in benchmark datasets) distort model ranking~\cite{shojaee2025illusion,yao2024secondhalf,openai2023eval}, and benchmark scores frequently fail to predict practical utility~\cite{yao2024secondhalf,dehghani2021benchmark}. Research on these challenges focuses on \emph{what} evaluation should measure while treating the software infrastructure that executes evaluation as a transparent medium. Whether contamination in benchmark data is detected, reproducibility of results is enforced, or annotation quality of benchmark labels is validated depends in practice on the engineering of evaluation infrastructure.

\subsection{From Ad-Hoc Scripts to Evaluation Infrastructure}
\label{sec:bg_infrastructure}

The ML community has invested in evaluation infrastructure over time. Early evaluation relied on ad-hoc scripts and manual processes that were difficult to reproduce and prone to errors. Standardized benchmark suites such as GLUE~\cite{wang2018glue} for language understanding and ImageNet~\cite{deng2009imagenet} for vision established common evaluation protocols and enabled meaningful comparison across research groups. The emergence of foundation models accelerated this trend: projects such as HELM~\cite{liang2022holistic}, BigCode Eval~\cite{srivastava2023beyond}, and LM Eval~\cite{eval-harness} provide standardized interfaces for assessing models across multiple dimensions and use cases.

This infrastructure evolution reveals an architectural distinction often conflated in the literature. \textbf{Benchmarks} define the \emph{what} of evaluation: tasks, datasets, ground truth references, and scoring metrics that establish correctness criteria. \textbf{Evaluation harnesses} provide the \emph{how}: the software that operationalizes measurement through model invocation protocols, resource management, error handling, result aggregation, and reporting interfaces. Benchmark validity (whether a metric captures the intended construct) and operational reliability (whether infrastructure executes measurement correctly) are orthogonal engineering challenges. A theoretically sound metric implemented in fragile infrastructure yields unreliable results; conversely, operationally reliable infrastructure can surface methodological limitations through contamination checks, reproducibility enforcement, and annotation validation. Existing literature engages primarily with the benchmark side of this distinction; the following review examines how evaluation engineering remains underexplored across three relevant research areas.

\subsection{Related Work}
\label{sec:related_work}

\subsubsection{Evaluation Methodology Surveys}

A large body of survey work examines ML evaluation from the perspective of what properties of models to measure. Chang et al.~\cite{chang2024survey} and Zhao et al.~\cite{zhao2023survey} survey LLM evaluation across tasks, metrics, and benchmarks. Domain-specific surveys cover reasoning capabilities~\cite{xia2025evaluating,mondorf2024beyond}, bias detection~\cite{gallegos2024bias,ecurali2024automated}, robustness~\cite{cecchini2024holistic,zhang2025evaluating}, and security assessment~\cite{zhou2024survey}. These surveys catalog evaluation dimensions and identify methodological gaps, but, to our knowledge, none examine the software that executes evaluations. They treat evaluation harnesses as interchangeable tools rather than engineered software with its own operational characteristics, failure modes, and design tradeoffs.

\subsubsection{MLOps and SE for ML}

The MLOps literature addresses operational challenges in ML systems broadly. Sculley et al.~\cite{sculley2015hidden} identified technical debt in ML systems, noting that surrounding infrastructure introduces most maintenance burden. Amershi et al.~\cite{amershi2019software} studied SE practices at Microsoft and found that data management, model evolution, and deployment posed distinct engineering challenges compared to traditional software. Subsequent work has formalized ML pipeline stages covering data ingestion, feature engineering, training, and deployment~\cite{ashmore2021assuring,paleyes2022challenges,kreuzberger2023machine}.
Within this literature, evaluation appears as a pipeline stage (typically ``model validation'' or ``testing'') rather than an operational domain in its own right. As a result, most frameworks specify when evaluation occurs but offer limited guidance on how harnesses handle dependency volatility, execution failures, and result integrity in practice. MLOps frameworks treat evaluation as a checkpoint between training and deployment, not as an activity requiring its own workflows, infrastructure management, and failure mitigation.

\subsubsection{Software Testing Infrastructure}

Software testing research offers structural parallels to EvalEng. Test automation frameworks manage test selection, execution orchestration, result collection, and failure reporting~\cite{garousi2018smells}. Continuous integration systems~\cite{hilton2016usage,widder2019conceptual} address many of the same operational concerns: environment provisioning, dependency management, execution scheduling, and result persistence. Flaky test research~\cite{luo2014empirical,parry2021survey} studies non-determinism in test outcomes, a concern that parallels stochastic evaluation results in ML.

However, evaluation harnesses differ from traditional test infrastructure in several respects. Evaluation involves heterogeneous external dependencies (pre-trained models, benchmark datasets, and third-party APIs) that traditional test suites do not manage. Metrics in ML evaluation are often continuous and aggregate rather than binary pass/fail, which makes error detection less straightforward because infrastructure faults can appear as small score shifts rather than explicit test failures. Evaluation runs are computationally expensive, typically requiring GPU scheduling and distributed execution. These differences indicate that the SE testing principles apply partially but do not cover the full operational scope of ML evaluation.

\subsection{The Missing Operational Perspective}
\label{sec:gap}

Evaluation surveys focus on \emph{what} to measure, while the software that carries out the measurement receives little attention. MLOps research covers the ML lifecycle but treats evaluation as a pipeline checkpoint. Software testing research addresses execution infrastructure but not the domain-specific challenges of ML evaluation.

Evaluation engineering shares concerns with MLOps, such as dependency management, environment reproducibility, and pipeline orchestration, but diverge in several respects. First, evaluation harnesses integrate heterogeneous external artifacts (pre-trained models, benchmark datasets, third-party scoring APIs) that vary across evaluation runs, whereas MLOps pipelines typically operate on a fixed model and dataset per training job. Second, evaluation produces continuous, aggregate metrics rather than binary pass/fail verdicts, making silent scoring errors harder to detect. Third, evaluation harnesses increasingly rely on LLM-based judges for subjective assessment, introducing a dependency on external model behavior that has no parallel in traditional MLOps testing stages.

To our knowledge, no previous work has studied evaluation tools as software products with their own workflows, the challenges developers encounter, and the engineering decisions that shape their reliability. Our work addresses this gap through empirical analysis of $57$ evaluation harnesses and $19{,}638$ GitHub issues.

\section{Methodology}
\label{sec:methodology}

\begin{figure}[!t]
\centering
\includegraphics[width=\columnwidth]{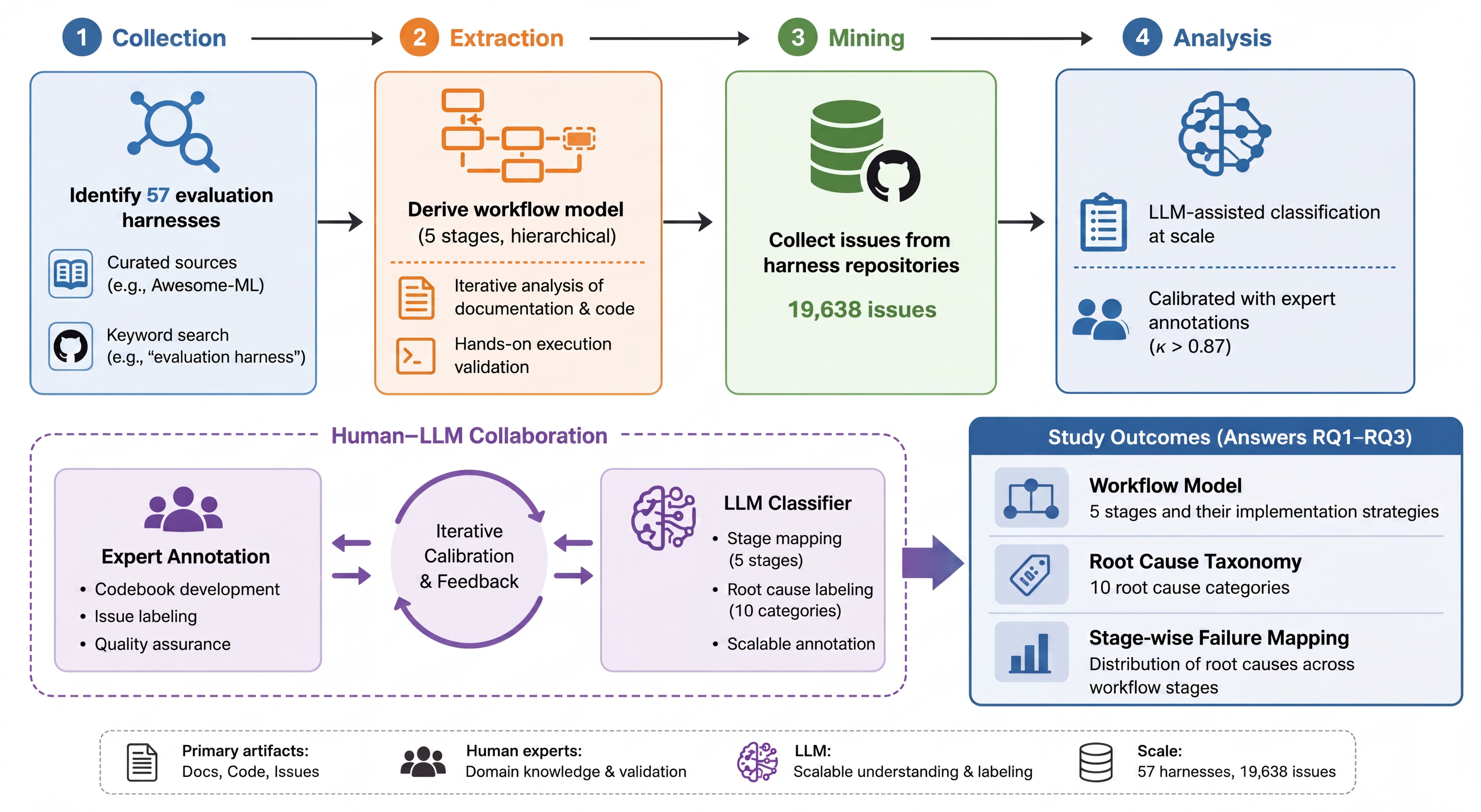}
\caption{Study workflow showing the four-stage methodology for investigating ML evaluation harnesses.}
\label{fig:study_workflow}
\end{figure}

Our methodology proceeds in four stages (Figure~\ref{fig:study_workflow}): (1) collect evaluation harnesses and their documentation (\S\ref{sec:method:harness_collection}); (2) extract evaluation workflows (\S\ref{sec:method:workflow_extraction}); (3) collect GitHub issues from the collected harnesses (\S\ref{sec:method:issue_collection}); and (4) analyze the issues to answer RQ2 and RQ3 (\S\ref{sec:method:issue_analysis}).

\subsection{Evaluation Harnesses and Documentation Collection}
\label{sec:method:harness_collection}

In our study, we define an \emph{evaluation harness} as a software framework whose primary purpose is to orchestrate ML model evaluation, as distinct from (1) benchmark repositories that provide only datasets without a configurable evaluation API, (2) standalone metric computation libraries whose sole purpose is providing scoring functions without model invocation or result orchestration, and (3) comprehensive ML frameworks that include evaluation as a single step in a broader training or deployment pipeline. We identify an initial set of harnesses (hereafter \emph{seed harnesses}) from curated sources, broaden coverage via keyword-based searches seeded by these harnesses' self-descriptions, and aggregate online documentation from multiple sources.

\subsubsection{Seed Harnesses Identification from Curated Sources}
\label{sec:method:seed_harnesses}

To ensure baseline quality, we start from the Awesome Production ML List\footnote{\url{https://github.com/EthicalML/awesome-production-machine-learning}}, a community-curated ML production resource list (20k+ GitHub stars, maintained since 2018). From its ``Evaluation and Monitoring'' section, the first two authors extract $45$ harnesses whose primary purpose is ML model evaluation, and contribute $12$ newly identified evaluation harnesses back to this list during the study. The keywords practitioners use to describe these seed harnesses inform the keyword-based search described next.

\subsubsection{Harness Coverage Expansion via Keyword-Based Search}
\label{sec:method:keyword_search}

We expand our harness collection through keyword-based GitHub searches~\cite{bhatia2023empirical}. For each seed harness, we examine its README file to extract self-described evaluation-related keywords (\eg, ``evaluation library'', ``benchmarking suite'') commonly used to characterize ML evaluation tools. By aggregating keywords across all seed harnesses, we identify a total of $25$ distinct keyword phrases (Table~\ref{tab:harness_search_keywords}). We then use each keyword phrase to conduct GitHub searches. For every retrieved repository, the first two authors independently verify whether its primary purpose aligns with ML model evaluation based on three criteria: (1) the repository's README explicitly describes evaluation or benchmarking as its core function, (2) the codebase implements model invocation, metric computation, or result reporting, and (3) the repository satisfies the inclusion criteria specified in the Awesome Production ML List's CONTRIBUTION guidelines (at least $500$ GitHub stars and evidence of activity within the past 12 months).

Table~\ref{tab:harness_search_keywords} presents search results showing both total retrieval counts and repositories meeting our criteria. Some keywords with high retrieval counts yield no qualifying harnesses for two reasons: the retrieved repositories may serve purposes outside ML model evaluation (\eg ``testing tool'' predominantly retrieves general software testing frameworks), or the matching repositories fall below the quality thresholds. This process yields $57$ evaluation harnesses spanning multiple ML domains (\eg, language modeling, computer vision, reinforcement learning, and general ML systems).

\subsection{Evaluation Workflow Extraction}
\label{sec:method:workflow_extraction}

Using iterative open card sorting with constant comparison~\cite{glaser1967strauss}, we analyze harness documentation, triangulate ambiguities through source code inspection and local execution, and consolidate the results into a hierarchical workflow model.

\subsubsection{Iterative Harnesses Documentation Analysis}
\label{sec:method:doc_analysis}

The first two authors independently perform open card sorting~\cite{spencer2009card} on the documentation of all $57$ harnesses, deriving workflow categories from the data rather than applying a predefined scheme. We prioritize the main README to reconstruct each evaluation workflow, consulting additional sources (\eg, GitHub Wiki, official website, technical report) as needed. When documentation is ambiguous, we triangulate through source code inspection or by running individual components locally in a clean Python environment (\eg, examining grading logic when the README lacks detail on supported metrics). We record \emph{operational steps}, defined as concrete user actions required to run an evaluation (\eg, installing dependencies), and use them to characterize the workflow. Our analysis reaches theoretical saturation~\cite{glaser1967strauss} (\ie, the point at which new data yield no new analytic categories) at the $51^{st}$ harness, after which no new step categories emerge from the remaining five harnesses.

\subsubsection{Evaluation Workflow Model Development}
\label{sec:method:taxonomy_development}

In these sessions, the first two authors apply continuous comparison~\cite{glaser1967strauss}: each operational action extracted from the documentation (\eg, generating a leaderboard) is compared against the emerging workflow model, either merging it into an existing category or creating a new one when no existing category fits. We organize the resulting categories into three hierarchical layers:
\begin{itemize}[leftmargin=*]
    \item \textbf{Stages} are high-level phases of the evaluation lifecycle that follow a logical progression (\eg, Provisioning $\rightarrow$ Execution $\rightarrow$ Reporting).
    \item \textbf{Steps} are distinct functional tasks within a stage (\eg, harness installation and credential configuration within Provisioning).
    \item \textbf{Strategies} are alternative technical implementations for accomplishing a step (\eg, git clone, Python package, or container image for harness installation).
\end{itemize}
We first identify concrete operational tasks (actions a user must perform to run an evaluation, such as installing dependencies or loading a dataset) as steps, then aggregate functionally related steps into stages and decompose each step into strategies when multiple implementation alternatives are observed.

Each author independently labels which stages, steps, and strategies each harness supports. We then cross-compare our labels and negotiate iteratively until reaching consensus for all $57$ harnesses. The final workflow model comprises $5$ sequential stages, $9$ operational steps, and $34$ implementation strategies. We encode the results in a $57 \times 9$ harness-step support matrix. Because certain steps are inapplicable to some harnesses (\eg, a scoring library that accepts pre-computed outputs has no SUT invocation step), $6.4\%$ ($33$) of cells in the matrix are naturally empty.

\subsubsection{Clustering Harnesses by Workflow Support Patterns}
\label{sec:method:harness_clustering}

We construct a binary feature matrix where cell $(i, j) = 1$ indicates harness $i$ supports strategy $j$, and apply Ward's hierarchical clustering~\cite{ward1963hierarchical}, which minimizes within-cluster variance at each merge step, to group harnesses with similar strategy coverage into evaluation archetypes. We select the dendrogram cut point by examining silhouette scores~\cite{rousseeuw1987silhouettes} across candidate cluster counts ($k = 2$--$8$), where $k = 6$ yields the highest mean silhouette score. The first two authors then jointly review the six cluster compositions and consolidate them into four evaluation archetypes based on two criteria: (1) shared workflow coverage patterns and domain focus (\eg, LLM-based evaluation) across cluster members, and (2) sufficient cluster size to avoid singleton or very small groupings. Appendix Figure~\ref{fig:rq1_pca} projects the resulting clusters onto the first two principal components via PCA~\cite{abdi2010principal}, confirming that the four archetypes occupy geometrically distinct regions in strategy space.

\subsection{GitHub Issues Collection}
\label{sec:method:issue_collection}

To investigate root causes (RQ2) and their distribution across the workflow (RQ3), we mine GitHub issue reports from our collected harness repositories. We retrieve both open and closed issues up to \collectdate{} to capture the complete spectrum of problems, from ongoing investigations to resolved issues with documented solutions. This process retrieves $19{,}638$ issues from $59$ GitHub repositories ($57$ harnesses, two of which maintain separate backend and frontend repositories).

\subsection{GitHub Issues Analysis}
\label{sec:method:issue_analysis}

We first establish a classification methodology combining manual annotation with LLM-based classification, then apply it in two passes: first to map issues onto workflow stages, steps, and strategies (\S\ref{sec:method:taxonomy_development}), filtering to \emph{workflow-relevant issues} (issues that affect evaluation operations, as opposed to general software maintenance or off-topic requests), then to categorize root causes (RQ2).

\subsubsection{Classification Methodology}
\label{sec:method:classification_methodology}

Since the full dataset is too large to label manually, we develop a hybrid methodology combining manual annotation with LLM-based classification, as follows:
\begin{enumerate}[leftmargin=*]
    \item \textbf{Manual examination.} To ensure statistical significance, we first randomly sample $377$ issues from our corpus. This sample size provides a $95\%$ confidence level and a $5\%$ margin of error, assuming maximum variability in the underlying population ($p=0.5$)~\cite{cochran1977sampling}, after which the first two authors independently annotate the sampled issues using a predefined taxonomy tailored to the specific RQ (\eg, workflow stages and steps for RQ2 and root causes for RQ3). We assess inter-rater reliability using Cohen's kappa ($\kappa$)~\cite{landis1977measurement} and resolve discrepancies through joint review to establish consensus.
    \item \textbf{LLM calibration.} We build a Claude Haiku 4.5-based classifier (``anthropic/claude-haiku-4.5''\footnote{https://www.anthropic.com/claude/haiku}, default configurations, 200K-token context window). The classification system comprises: (1) a system prompt with workflow definitions and annotation guidelines, and (2) a user prompt containing the full issue context (title, body, and comments), truncated at 200K tokens when the issue exceeds the context window. We iteratively refine the prompt to address edge cases until achieving substantial agreement ($\kappa > 0.8$) between LLM classifications and human consensus labels.
    \item \textbf{Large-scale annotation.} Finally, we apply the calibrated LLM classifier to annotate the remaining issues.
\end{enumerate}
All intermediate annotations, classifier prompts, and final classification labels are available in the replication package~\cite{replication_package}.

\subsubsection{Workflow Classification}
\label{sec:method:workflow_annotation}

We apply closed card sorting (classifying against the workflow model established above) to map issues onto the workflow model (\S\ref{sec:method:taxonomy_development}), extracting: (1) workflow relevance, whether the issue affects evaluation operations; (2) workflow stage; (3) operational step; and (4) implementation strategy if identifiable. When issues affect multiple components, we assign the primary label based on the most direct operational impact. This classification serves two purposes: it identifies the $16{,}560$ workflow-relevant issues ($84.3\%$) that form the corpus for root cause analysis (RQ2), and it provides the stage/step/strategy labels needed to examine where root causes concentrate across the workflow (RQ3). Our manual annotation yields Cohen's kappa ($\kappa = 0.894$, $91.8\%$ raw agreement), indicating substantial inter-rater reliability, and LLM validation achieves Cohen's kappa ($\kappa = 0.931$, $94.2\%$ raw agreement) against human consensus. Of the $377$ sampled issues, $50$ are non-workflow-relevant and excluded from subsequent root cause analysis, leaving $327$ workflow-relevant issues.

\subsubsection{Root Cause Classification and RQ-Specific Cross-Tabulation}
\label{sec:method:root_cause_annotation}

For root cause classification, the first two authors examine the $327$ workflow-relevant sample issues to identify common failure categories through open card sorting. After three negotiation rounds, the authors reach consensus on a final taxonomy of ten root causes. Inter-rater agreement reaches Cohen's kappa ($\kappa = 0.758$, $78\%$ raw agreement), with one issue not fitting any category. The lower $\kappa$ relative to workflow classification ($0.894$) reflects the inherent ambiguity of root cause attribution, as a single issue can plausibly involve multiple interacting causes. After LLM calibration against consensus labels (Cohen's kappa $\kappa = 0.873$, $89.3\%$ raw agreement), we classify the full corpus, and $207$ issues ($1.3\%$) fall outside the ten categories. To answer RQ2, we cross-tabulate each issue's root cause label with the harness archetype assigned to its repository (\S\ref{sec:method:harness_clustering}), producing archetype-specific root cause distributions. To answer RQ3, we cross-tabulate each issue's root cause label with its workflow stage label (from \S\ref{sec:method:workflow_annotation}), producing stage-specific root cause distributions.

To illustrate the classification process, consider issue \#1407 from LM Eval\footnotemark[14], titled ``Multi-GPU evaluation fails with AssertionError''\footnote{\url{https://github.com/EleutherAI/lm-evaluation-harness/issues/1407}}. A user reports that running evaluation across multiple GPUs triggers an assertion failure during model loading. \emph{Workflow classification:} the issue affects how the model is loaded onto hardware, which falls under the Specification stage (S1), SUT preparation step (S1-A), model-in-process strategy (S1-A1). \emph{Root cause classification:} the harness lacks logic for distributing model weights across devices, so we label it as \emph{unimplemented feature gap}. This example shows how a single issue receives both a workflow label (where it occurs) and a root cause label (why it occurs).

\section{RQ1: Unified Workflow for Evaluation Harnesses}
\label{sec:rq1-results}

\subsection{Workflow Component Definition}

\begin{figure}[!t]
\centering
\includegraphics[width=\columnwidth]{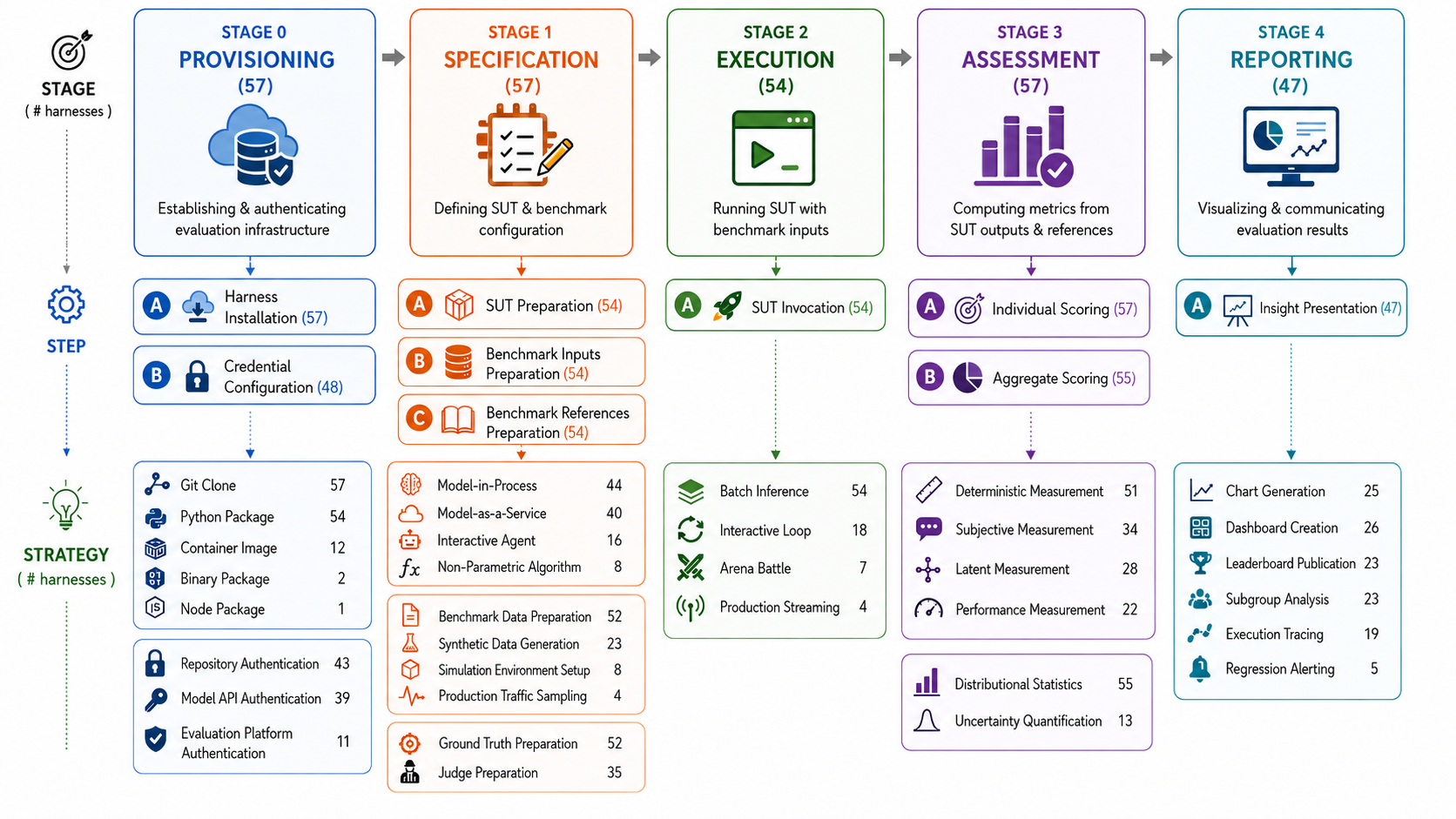}
\caption{Operational workflow for evaluation harnesses, depicting a five-stage lifecycle from provisioning through reporting. Parenthesized numbers indicate how many harnesses support each stage, step, or strategy.}
\label{fig:rq1_workflow}
\end{figure}

The operational workflow for evaluation harnesses follows a five-stage progression (Figure~\ref{fig:rq1_workflow}), starting from establishing the runtime environment (Provisioning) and defining evaluation contracts (Specification), through executing the System-under-Test (SUT), the model or system being evaluated, (Execution), to quantitatively measuring execution outcomes (Assessment) and finally producing actionable insights (Reporting) for stakeholders. Appendix Table~\ref{tab:workflow-components} provides the full definitions and examples for all workflow components.

\subsection{Workflow Component Analysis}

\begin{figure}[!t]
\centering
\includegraphics[width=\columnwidth]{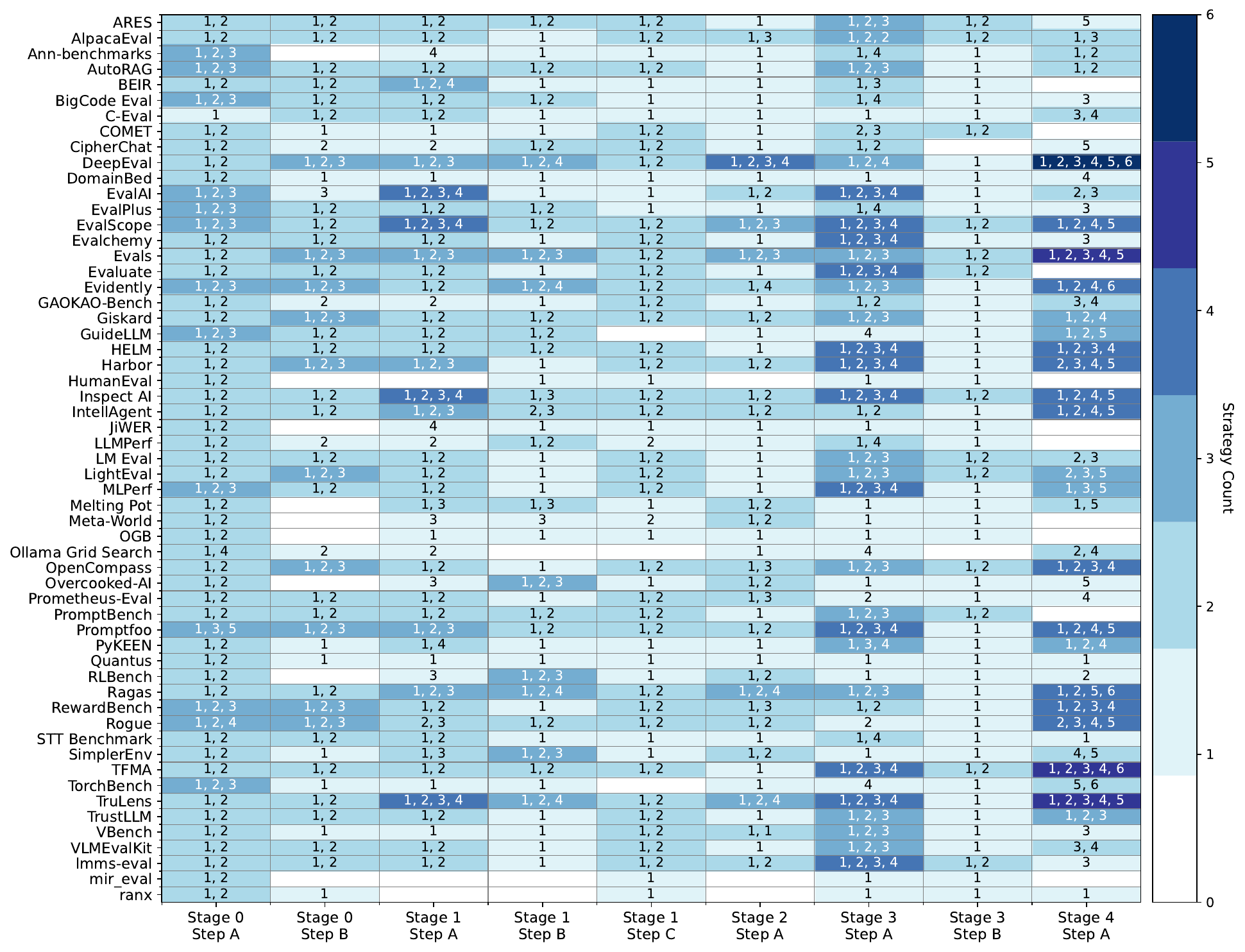}
\caption{Strategy support heatmap across $57$ evaluation harnesses (rows) and $9$ workflow steps (columns). Cell intensity indicates how many of the step's strategies each harness implements (white: 0, light orange: 1, orange: 2--4, dark red: 5--6).}
\label{fig:rq1_heatmap}
\end{figure}

Figure~\ref{fig:rq1_heatmap} presents the complete strategy support matrix across all harnesses, showing which specific strategies each harness supports at each workflow step.

\subsubsection{S0: Provisioning}

\textbf{Containerized deployment is rarely supported despite all harnesses supporting source-based installation.} While git clone (S0-A1, 100\%) and Python package (S0-A2, 94.7\%) are adopted by nearly all harnesses, container image (S0-A3) adoption reaches only 21.1\% (12 of 57 harnesses). This means developers who need reproducible, isolated environments must build and maintain container configurations independently rather than relying on pre-built images provided by the harness.

\textbf{Credential configuration centers on model and dataset access rather than on platform integration.} Most harnesses require both repository authentication (S0-B1, 75.4\%) to retrieve models and datasets from repositories, and model API authentication (S0-B2, 68.4\%) to access model-serving endpoints. In contrast, evaluation platform authentication (S0-B3) reaches only 19.3\%, reflecting that most harnesses operate as local tools that produce results independently rather than relying on external platform services.

\subsubsection{S1: Specification}

\textbf{Offline, reference-based evaluation setup is the most prevalent.} Offline benchmark inputs and ground-truth references are adopted by 91.2\% of harnesses (S1-B1 and S1-C1), indicating that most harnesses define evaluation around predefined inputs and reference targets. Interactive agent evaluation is supported by a minority of harnesses (S1-A3, 28.1\%), indicating that evaluation setups requiring multi-step interaction are less commonly covered at this stage.

\textbf{Production-traffic inputs represent the largest gap in harness specification.} Production traffic sampling (S1-B4), which enables evaluation on real-world user inputs, appears in only 4 of 57 harnesses (7.0\%), whereas offline dataset loading is supported by 91.2\% of harnesses (S1-B1). This gap suggests that users evaluating on production traffic often require external traffic capture and replay, rather than relying on built-in harness support.

\textbf{Reference-based scoring sees higher adoption than judge-based scoring.} Judge preparation (S1-C2, 61.4\%) is less prevalent than ground truth preparation (S1-C1, 91.2\%), indicating that many harnesses still operationalize evaluation primarily through reference targets instead of configured judges. Consequently, the judge configuration is often handled outside the harness, which can hinder end-to-end reproducibility.

\subsubsection{S2: Execution}

\textbf{Execution centers on a single strategy: batch inference.} Nearly all harnesses support batch inference (S2-A1, 94.7\%), processing multiple inputs through a single SUT instance in one pass. The only alternative with notable adoption is interactive loop (S2-A2, 31.6\%), which enables stateful, multi-turn agent evaluation. The remaining strategies, arena battle (S2-A3, running multiple SUTs on the same input for pairwise comparison) and production streaming (S2-A4, continuously processing live inference traffic), are adopted by only 12.3\% (7 harnesses) and 7.0\% (4 harnesses) respectively, consistent with the low adoption of production traffic sampling (S1-B4, 7.0\%) in Stage~1. These adaptation patterns suggest that most harnesses are designed primarily for static input-output testing rather than dynamic or online evaluation.

\subsubsection{S3: Assessment}

\textbf{Assessment scoring relies heavily on deterministic metrics.} Deterministic measurement (S3-A1) is supported by 89.5\% of harnesses, while the remaining scoring strategies range from 59.6\% to 38.6\% adoption. This declining adoption indicates that as scoring moves from exact-match-style metrics toward judgment-based, embedding-based, or efficiency-based evaluation, harness support narrows considerably.

\textbf{Harnesses compute aggregate scores but rarely quantify their statistical confidence.} Distributional statistics such as means and weighted aggregates (S3-B1) are supported by 96.5\% of harnesses. However, only 22.8\% of harnesses support uncertainty quantification (S3-B2), meaning most harnesses cannot indicate whether an observed score difference is meaningful or due to chance.

\subsubsection{S4: Reporting}

\textbf{Reporting is the least supported stage in the workflow.} Unlike earlier stages where at least one strategy exceeds 89\% adoption, no reporting strategy surpasses 45.6\% (dashboard creation, S4-A2), suggesting that most harnesses treat visualization and result presentation as optional. Regression alerting (S4-A6, 8.8\%) is the least adopted strategy across the entire workflow, indicating that harnesses largely lack the ability to automatically flag performance degradation between runs.

\subsection{Harness Archetype Definition and Analysis}

Table~\ref{tab:archetypes} summarizes the four archetypes, their prevalence, and defining workflow strategies.

\begin{table}[!t]
\centering
\small
\caption{Four evaluation archetypes identified through hierarchical clustering of strategy support patterns across harnesses.}
\label{tab:archetypes}
\begin{tabularx}{\columnwidth}{p{2.0cm}Xp{1.7cm}p{1.7cm}}
\textbf{Archetype (\%)} & \textbf{Definition} & \textbf{Common Strategies} & \textbf{Missing Strategies} \\
\midrule

Standardized LLM Benchmark Suites (40.4\%) &
Batch inference harnesses that sweep foundation models across fixed, published task collections and produce normalized leaderboard scores via embedding-based or model-judged metrics &
S0-A1\newline S1-B1, S1-C1\newline S2-A1\newline S3-B1 &
S0-A4, S0-A5\newline S1-A3, S1-A4, S1-B3\newline S2-A4 \\
\midrule

Narrow-Domain Metric Libraries (21.1\%) &
Single-metric libraries that score structured inputs locally without invoking any remote model API, LLM judge, or production plumbing &
S0-A1, S0-A2\newline S3-A1, S3-B1 &
S0-A3--S0-A5\newline S0-B2, S0-B3\newline S1-A2, S1-B4\newline S2-A3, S2-A4\newline S3-A2--S3-A4, S3-B2\newline S4-A3, S4-A6 \\
\midrule

Task-Specific Capability Probes (21.1\%) &
Targeted probes for a single capability axis (retrieval, code correctness, inference latency, or adversarial safety) via remote model invocation and scalar leaderboard scoring &
S0-A1\newline S2-A1 &
S0-A5\newline S1-B3, S1-B4\newline S2-A3, S2-A4\newline S3-B2 \\
\midrule

Full-Stack LLM Evaluation Platforms (17.5\%) &
Persistent evaluation infrastructure that spans all execution modes (batch regression, arena head-to-head, agentic loop tracing, and production monitoring), with every strategy present in at least one member &
S0-A1, S0-B1, S0-B2\newline S1-A2, S1-A3, S1-C1, S1-C2\newline S2-A1, S2-A2\newline S3-A2, S3-B1\newline S4-A2, S4-A5 &
None \\
\end{tabularx}
\end{table}

\textbf{Standardized LLM Benchmark Suites (40.4\%) and Narrow-Domain Metric Libraries (21.1\%) together cover over 61\% of harnesses, both restricted to static, offline workflows.} Standardized LLM Benchmark Suites sweep foundation models across fixed, published task collections to produce leaderboard scores, sharing five common strategies (S0-A1, S1-B1, S1-C1, S2-A1, S3-B1) while consistently omitting interactive agent evaluation (S1-A3, S1-A4) and real-time inference monitoring (S1-B3). Narrow-Domain Metric Libraries compute one metric over locally available inputs without invoking any remote model or judge; with only four common strategies and 15 missing (27.5\% average coverage, the lowest of any archetype), each harness covers only what its single scoring function requires.

\textbf{Task-Specific Capability Probes and Full-Stack LLM Evaluation Platforms both invoke remote models.} Task-Specific Capability Probes (21.1\%) target a single capability axis (retrieval precision, code correctness, inference latency, or adversarial safety) and always invoke a remote endpoint (S0-A1 and S2-A1 adopted by all probes), reaching 37.4\% average coverage, but still omit judge-based scoring (S1-C2), persistent dashboards (S4-A2), and production monitoring (S2-A4). Full-Stack LLM Evaluation Platforms (17.5\%) are the only archetype where every strategy is adopted by at least one member (69.7\% average coverage), meaning a single platform can serve any evaluation scenario. They are the only harnesses that simultaneously support batch regression, arena head-to-head comparison (S2-A3, running multiple SUTs on the same input for pairwise comparison), agentic loop tracing (S2-A2), and production monitoring (S2-A4), and the only archetype where judge preparation (S1-C2) and subjective measurement (S3-A2) are consistently supported, reflecting LLM-as-judge~\cite{zheng2023judging} as a first-class evaluation mode.

\begin{finding}{Summary (RQ1): Harnesses strongly support offline execution but lag in reporting and online evaluation}
\begin{itemize}[leftmargin=*]
    \item \textbf{Regression alerting is the least adopted (8.8\%) strategy in the workflow.} Most harnesses have no automated mechanism for detecting performance degradation across evaluation runs, a capability that SE practice has long addressed through both commit-time quality gates and post-deployment canary testing.
    \item \textbf{Three of the four archetypes concentrate on offline, batch evaluation, while only Full-Stack LLM Evaluation Platforms span all five workflow stages.} Standardized LLM Benchmark Suites (40.4\%) and Narrow-Domain Metric Libraries (21.1\%) are both restricted to static, offline workflows. Task-Specific Capability Probes (21.1\%) extend to online evaluation via remote model API invocation but still omit production traffic monitoring and judge-based scoring. Only Full-Stack LLM Evaluation Platforms (17.5\%) span all five stages with every strategy present in at least one member.
\end{itemize}
\end{finding}

\section{RQ2: Root Causes of Operational Challenges}
\label{sec:rq2-results}

\subsection{Root Cause Definition \& Prevalence}

Table~\ref{tab:root_cause_definitions} presents the ten root cause categories, each with its definition and the percentage of issues attributed to it.

\begin{table}[!t]
\centering
\caption{Ten root cause categories for operational challenges in ML evaluation harnesses, with the percentage of issues attributed to each category. Categories cover both defects in existing functionality and capability gaps that block intended workflows.}
\label{tab:root_cause_definitions}
\begin{tabularx}{\columnwidth}{
    >{\raggedright\arraybackslash}p{0.18\columnwidth}
    r
    >{\raggedright\arraybackslash}X
}
\textbf{Root Cause} & \textbf{\%} & \textbf{Definition} \\
\midrule
Unimplemented Feature Gap & 24.26 & Required functionality is not implemented, leaving expected capabilities unavailable. \\
\midrule
Documentation Deficiency & 20.27 & Documentation is missing, incomplete, or outdated, so users cannot correctly use implemented functionality. \\
\midrule
Validation Gap & 17.17 & Input, output, or state validation is missing or insufficient, allowing invalid conditions and weak error handling. \\
\midrule
Algorithmic Error & 8.27 & Code executes but produces incorrect results due to flaws in metric implementations, scoring functions, or aggregation logic. \\
\midrule
External Dependency Breakage & 7.56 & Changes or outages in third-party libraries, APIs, or services break previously working behavior. \\
\midrule
Configuration Error & 6.95 & Configuration mechanisms exist, but values fail to propagate correctly, are missing, or use inappropriate defaults. \\
\midrule
Environment Incompatibility & 5.19 & The system assumes specific platforms, Python versions, or hardware, causing failures in other environments. \\
\midrule
Architectural Constraint & 3.26 & Core design choices block required adaptation or extension, so fixes require refactoring rather than localized patches. \\
\midrule
Interface Contract Mismatch & 2.95 & Integrated components disagree on data types, formats, or API signatures at their boundaries. \\
\midrule
Resource Mishandling & 2.86 & Memory, GPU resources, file handles, connections, or concurrency primitives are allocated, used, or released incorrectly. \\
\end{tabularx}
\end{table}

\textbf{Challenges mostly correspond to capability and documentation gaps rather than low-level runtime issues.} \textit{Unimplemented feature gap} (24.3\%), \textit{documentation deficiency} (20.3\%), and \textit{validation gap} (17.2\%) together account for 61.7\% of issues. By contrast, \textit{interface contract mismatch}, \textit{resource mishandling}, and \textit{architectural constraint} together account for only 9.1\%.

\subsection{Root Cause Distribution Across Archetypes}

\begin{figure}[!t]
\centering
\includegraphics[width=\columnwidth]{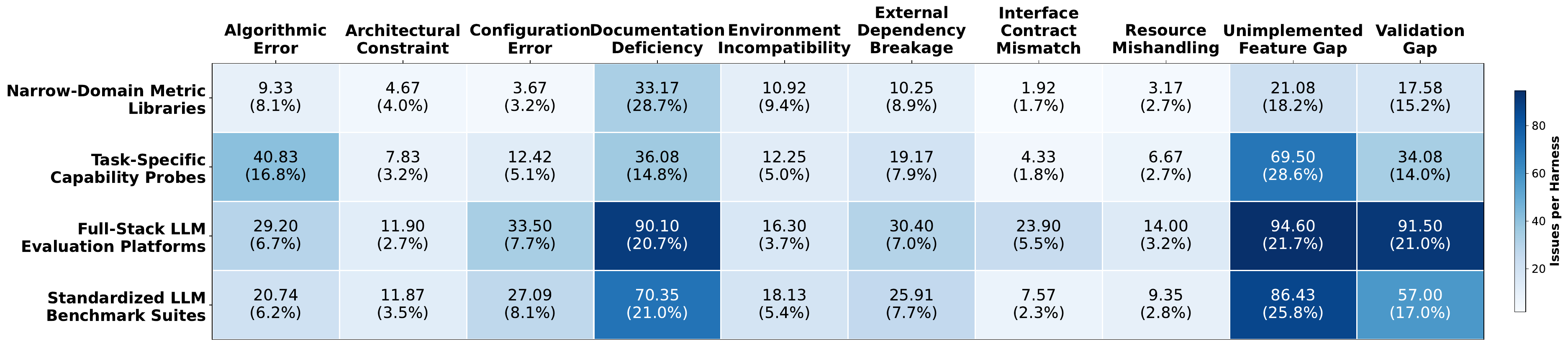}
\caption{Root cause distribution across the four harness archetypes. Rows represent the harness archetypes and columns represent the root cause categories (Table~\ref{tab:root_cause_definitions}). Cell color intensity scales with the normalized issue count (total issues divided by the number of harnesses in the archetype), from light (low) to dark (high).}
\label{fig:root_cause_archetype}
\end{figure}

Figure~\ref{fig:root_cause_archetype} cross-tabulates the ten root causes against the four harness archetypes. Each cell reports two values: a \emph{normalized issue count} (total issues divided by the number of harnesses in the archetype) for comparing issue volume across archetypes of different sizes, and a \emph{within-archetype percentage} (the root cause's share of all issues in that archetype, so each row sums to $100\%$) for comparing how each archetype distributes its issues across root causes.

\textbf{Full-Stack LLM Evaluation Platforms and Standardized LLM Benchmark Suites accumulate the highest per-harness issue volume, while Narrow-Domain Metric Libraries have the lowest.} When normalized by the number of harnesses in each archetype, Full-Stack platforms reach 93.3 issues per harness for \textit{unimplemented feature gap}, 90.6 for \textit{validation gap}, and 89.9 for \textit{documentation deficiency}; Standardized LLM Benchmark Suites reach comparable counts of 86.4, 57.0, and 70.3 respectively. Both archetypes dwarf Narrow-Domain Metric Libraries, which peak at 33.2 for \textit{documentation deficiency}, consistent with their single-metric scope that limits the number of components that can fail.

\textbf{Unimplemented feature gap leads in three archetypes, but each archetype has a distinct secondary root cause shaped by its operational focus.} Standardized LLM Benchmark Suites, Task-Specific Capability Probes, and Full-Stack LLM Evaluation Platforms all rank \textit{unimplemented feature gap} first (27.8\%, 42.9\%, and 17.6\% of within-archetype issues, respectively), meaning harness users consistently request capabilities the harness does not yet implement. Secondary root causes diverge along archetype boundaries. Standardized LLM Benchmark Suites depend on external datasets and packages for multi-task coverage, so \textit{external dependency breakage} (8.9\%) is their most prominent secondary issue: in lm-evaluation-harness\footnotemark[14], benchmark datasets such as \texttt{pile\_freelaw} become unavailable when Hugging Face Hub configurations change upstream\footnote{\url{https://github.com/EleutherAI/lm-evaluation-harness/issues/1714}}, breaking evaluation without any local code change. Task-Specific Capability Probes implement domain-tailored scoring algorithms, making \textit{algorithmic error} their secondary root cause at 11.9\% (the highest rate across all archetypes), because domain-tailored scoring functions have fewer reference implementations to validate against. Full-Stack LLM Evaluation Platforms coordinate remote API calls and LLM-as-judge pipelines across multiple service boundaries, so \textit{interface contract mismatch} reaches 10.6\% (vs.\ 0--9.1\% in other archetypes), reflecting the cost of multi-component integration~\cite{zheng2023judging}. Narrow-Domain Metric Libraries are the exception: \textit{documentation deficiency} (31.8\%) is their top root cause, overtaking \textit{unimplemented feature gap} (22.7\%), because a narrow interface exposes few functionality gaps but demands precise setup instructions to apply its metric correctly.

\begin{finding}{Summary (RQ2): Most challenges stem from capability gaps and missing documentation rather than low-level runtime issues, and each archetype surfaces distinct secondary root causes}
\begin{itemize}[leftmargin=*]
    \item \textbf{61.7\% of issues stem from capability and documentation gaps, not interface, resource, or architectural issues.} The top three root causes are \textit{unimplemented feature gap} (24.3\%), \textit{documentation deficiency} (20.3\%), and \textit{validation gap} (17.2\%). This distribution mirrors the pattern observed in maturing open-source projects, where feature requests and documentation gaps dominate early issue trackers before reliability and interface concerns become prominent~\cite{mens2008introduction}.
    \item \textbf{Unimplemented feature gap ranks first in three of four archetypes, but each archetype has a distinct secondary root cause.} Secondary root causes are \textit{external dependency breakage} in Standardized LLM Benchmark Suites, \textit{algorithmic error} in Task-Specific Capability Probes, and \textit{interface contract mismatch} in Full-Stack LLM Evaluation Platforms. Narrow-Domain Metric Libraries are the exception, where \textit{documentation deficiency} (31.8\%) overtakes \textit{unimplemented feature gap} (22.7\%).
\end{itemize}
\end{finding}

\section{RQ3: Root Causes across Workflow Stages}
\label{sec:rq3-results}

\begin{figure}[!t]
  \centering
  \includegraphics[width=\columnwidth]{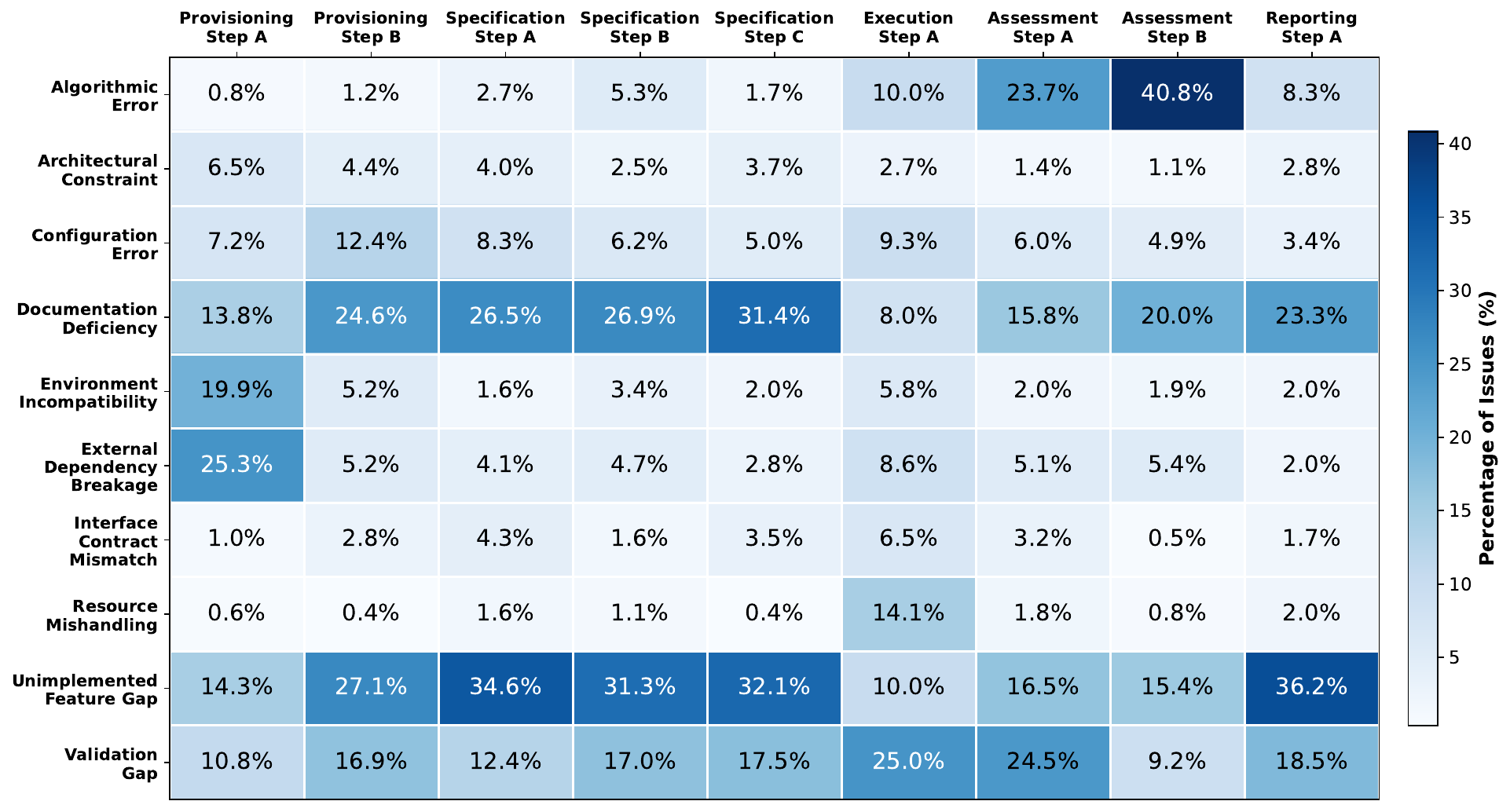}
  \caption{Root cause distribution across workflow steps. Cell color intensity scales with the within-root-cause percentage of issues, from light (low) to dark (high).}
  \label{fig:rq3_root_cause}
\end{figure}

Figure~\ref{fig:rq3_root_cause} cross-tabulates ten root causes against nine workflow steps. Each cell reports the \emph{within-root-cause percentage}, that is, the share of issues assigned to a given step within a root cause. The figure supports two complementary readings. On the one hand, the \emph{per-root-cause view} (reading across a row) shows how one root cause distributes across steps, yielding a \emph{within-root-cause share} for each step: for example, \textit{algorithmic error} places 43.3\% of its issues at individual scoring (S3-A) and 11.0\% at aggregate scoring (S3-B). Since both steps belong to the Assessment stage, summing them yields a stage-level assessment share of $43.3\% + 11.0\% = 54.3\%$. On the other hand, the \emph{per-step/stage view} (reading down a column or stage-level column group) shows which root causes dominate a given step or stage, yielding a \emph{step/stage-level share} computed by dividing each root cause's issue count at that step or stage by the corresponding total.

\subsection{Per-Root-Cause View: Distribution across Stages}

\textbf{Nearly all root causes concentrate over 50\% of their issues in one or two stages, while validation gap stays above 10\% across all five.} Using \emph{within-root-cause share}, \textit{external dependency breakage} (48.3\%) and \textit{environment incompatibility} (56.5\%) concentrate in provisioning, while \textit{unimplemented feature gap} (55.8\%) and \textit{documentation deficiency} (55.2\%) concentrate in specification, dropping to 5.1\% and 4.9\% at execution. The \textit{resource mishandling} root cause reaches 61.3\% at execution and the \textit{algorithmic error} root cause peaks at assessment (54.3\%), so these two root causes cluster at opposite ends of the workflow: resource issues arise when the SUT runs, and scoring issues surface when results are computed. A representative assessment-stage example is the \texttt{mean\_iou} metric in Hugging Face Evaluate\footnotemark[26], which computes recall instead of Intersection over Union because its denominator counts only true positives and false negatives (i.e., the ground-truth set), omitting false positives. As such, this formula actually matches the recall formula rather than the IoU formula\footnote{\url{https://github.com/huggingface/evaluate/issues/421}}, returning plausible scores without any runtime error until a user independently validated the output against a reference implementation. \textit{Validation gap}, by contrast, stays above 10\% across all five stages and peaks at assessment (22.7\%), making it the only cross-cutting root cause rather than a single-stage concern.

\textbf{Root causes vary substantially in how narrowly they localize to individual steps.} \textit{Resource mishandling} peaks at SUT invocation (S2-A: 61.3\%), \textit{environment incompatibility} and \textit{external dependency breakage} both peak at harness installation (S0-A: 51.9\% and 45.1\%), and \textit{algorithmic error} peaks at individual scoring (S3-A: 43.3\%), so each of these root causes has a clear step-level target. \textit{Interface contract mismatch}, by contrast, is nearly evenly split between SUT preparation and invocation (S1-A: 27.3\%, S2-A: 27.3\%), and \textit{configuration error} and \textit{validation gap} spread across steps with no single step exceeding 22.4\% and 21.5\%, respectively. In LM Eval\footnotemark[14], for instance, TriviaQA scores diverge dramatically between the \texttt{vllm} and \texttt{hf} backends ($0.615$ vs.\ $0.070$ for Llama-2-7B) because the two backends apply incompatible tokenization and generation contracts\footnote{\url{https://github.com/EleutherAI/lm-evaluation-harness/issues/1262}}, an interface contract mismatch that surfaces at both SUT preparation and invocation steps.

\subsection{Per-Stage View: Root-Cause Composition}

\textbf{Root-cause concentration increases from provisioning to specification and assessment, giving later stages more identifiable targets for improvement.} Using \emph{stage-level share}, provisioning spreads issues across four root causes with no single one exceeding 21\% (\textit{external dependency breakage} 20.1\%, \textit{unimplemented feature gap} 17.6\%, \textit{documentation deficiency} 16.6\%, \textit{environment incompatibility} 16.1\%). Specification and assessment each concentrate nearly half or more of their issues in two root causes: \textit{unimplemented feature gap} (32.8\%) and \textit{documentation deficiency} (27.1\%) account for 59.9\% of specification issues, while \textit{algorithmic error} (25.9\%) and \textit{validation gap} (22.5\%) account for 48.4\% of assessment issues. Execution and reporting fall between these extremes, with no single root cause exceeding 25\% in either stage. Specification and assessment challenges can therefore be traced to specific root cause pairs, whereas provisioning challenges arise from a broader mix of environmental and dependency factors. For instance, BigCode Eval\footnotemark[23] fails at installation due to conflicting \texttt{transformers} version requirements\footnote{\url{https://github.com/bigcode-project/bigcode-evaluation-harness/issues/141}}, a provisioning failure that combines \textit{external dependency breakage} and \textit{environment incompatibility} without a dominant single cause.

\textbf{Operational challenges shift from environment-related in early stages to scoring-related in later stages.} \textit{Environment incompatibility} (16.1\%) and \textit{external dependency breakage} (20.1\%) together account for 36.2\% of provisioning issues but drop below 6\% each by assessment. \textit{Algorithmic error}, conversely, accounts for only 0.9\% of provisioning issues but rises to 25.9\% in assessment. This shift is illustrated at the assessment stage, where harness users of OpenCompass\footnotemark[10] and lm-evaluation-harness\footnotemark[14] report different MMLU scores for the same model because the two harnesses use divergent prompt templates and scoring logic\footnote{\url{https://github.com/open-compass/opencompass/issues/466}}, a discrepancy that surfaces only when results are cross-compared across harnesses. Not all root causes follow this shift: \textit{unimplemented feature gap} and \textit{documentation deficiency} maintain notable shares in both specification (59.9\% combined) and reporting (59.5\% combined), persisting across early and late stages. The overall pattern suggests that early stages are constrained by the external environment a harness must integrate with, while later stages are constrained by the correctness of the harness's own scoring and validation logic.

\begin{finding}{Summary (RQ3): Within-stage root cause compositions shift from environment-related in early stages to scoring-related in later stages, with concentration increasing along the workflow}
  \begin{itemize}[leftmargin=*]
    \item \textbf{Nearly all root causes concentrate over 50\% of their issues in one or two stages, while validation gap stays above 10\% across all five.} For example, \textit{environment incompatibility} places 56.5\% in provisioning and \textit{algorithmic error} places 54.3\% in assessment. This stage-specific localization suggests that different failure modes are intrinsic to different workflow stages rather than randomly distributed across the pipeline.
    \item \textbf{Each stage has a distinct root-cause composition, and the challenges shift from environment-related in early stages to scoring-related in later stages.} Specification concentrates 59.9\% of its issues in \textit{unimplemented feature gap} and \textit{documentation deficiency}, while assessment is led by \textit{algorithmic error} (25.9\%) and \textit{validation gap} (22.5\%).
  \end{itemize}
\end{finding}

\section{Implications}
\label{sec:implications}

We discuss implications for the three communities engaged with evaluation infrastructure: harness developers who build and maintain these harnesses, harness users who depend on them for model assessment, and researchers who study evaluation as an object of inquiry.

\subsection{Implications for Harness Developers}

\emph{Enforce semantic API contracts across stages, not just schema checks.} RQ3 shows that \textit{validation gap} is the only root cause above $10\%$ in all five stages and peaks at assessment ($22.7\%$). This happens since evaluation harness stages exchange structured data (\eg, nested dictionaries, label ontologies) where data may be syntactically valid yet semantically incompatible with the downstream component. This goes against the recommendations of design-by-contract~\cite{meyer1992applying}, a technique that specifies software behavior through preconditions and postconditions at component boundaries to prevent such mismatches. For example, in COMET\footnotemark[22], the \texttt{layer\_transformation} configuration field accepts \texttt{sparsemax} as a valid value and passes schema validation, but two model subclasses (\texttt{UnifiedMetric} and \texttt{XCOMETMetric}) fail to forward the field to their base class, which silently defaults to softmax\footnote{\url{https://github.com/Unbabel/COMET/issues/195}}. The field is present with the correct type and a valid value, yet the downstream component applies a different activation function than specified, producing scores that deviate from the documented model behavior without any schema violation or runtime error. To avoid such silent failures, harness developers should adopt contract-based approaches that operate at the semantic level, encoding task-specific compatibility constraints (\eg, label vocabulary alignment, output modality matching) rather than relying on type or schema validation alone.

\emph{Build oracle-independent verification into scoring pipelines.} RQ3 shows that \textit{algorithmic error} concentrates $54.3\%$ of its issues in the Assessment stage ($43.3\%$ at individual scoring, $11.0\%$ at aggregate scoring), and these failures are characteristically silent: the harness produces plausible output without throwing any exception, so the defect escapes normal testing. For instance, LM Eval\footnotemark[14] reported ROUGE-L scores near $1.0$ for LLaMA-3.1-8B across all LongBench summarization tasks due to a metric computation bug\footnote{\url{https://github.com/EleutherAI/LM Eval/issues/2890}}, discovered only when a user cross-referenced the output against published paper results. This is a manifestation of the test oracle problem~\cite{barr2015oracle}: without an independent reference, the harness's own output becomes the implicit ground truth. Harness developers should therefore treat metric implementations as software under test. For example, metamorphic testing~\cite{chen2018metamorphic} encodes input--output invariants as regression cases associated with each newly introduced metric (\eg, a perfect candidate must not decrease a similarity score). Differential testing~\cite{mckeeman1998differential} cross-runs independent metric implementations as a release gate, catching defects that unit tests miss by returning consistent but wrong values.

\emph{Tailor maintenance priorities to archetype-specific failure modes rather than applying a uniform strategy.} RQ2 shows that the secondary root cause differs by archetype, so a one-size-fits-all backlog policy misallocates effort. For Standardized LLM Benchmark Suites, developers should pin dataset and package versions in a lockfile~\cite{he2025pinning} and add an import-time canary test that fails fast when upstream assets change, so the silent \textit{external dependency breakage} that is the archetype's top secondary cause ($8.9\%$) is caught before corrupting leaderboard scores. For Task-Specific Capability Probes, developers should encode score-monotonicity and boundary invariants as metamorphic regression cases~\cite{zhang2018deeproad} that execute on every commit, and run differential testing~\cite{mckeeman1998differential} against any available independent implementation before releasing the metric, since \textit{algorithmic error} reaches its highest rate across all archetypes ($11.9\%$) here precisely because domain-tailored scoring functions have no reference implementations to cross-validate against. For Full-Stack LLM Evaluation Platforms, developers should add contract tests~\cite{ayas2022contracttesting} at each service boundary that assert the judge's input schema matches what the scorer emits and that output formats remain stable across independently versioned services, targeting the \textit{interface contract mismatch} ($10.6\%$) that concentrates where LLM-as-judge pipelines and remote API calls must be coordinated. For Narrow-Domain Metric Libraries, developers should treat parameter docstrings as testable specifications~\cite{hossain2025doc2oracll} by pairing each with a worked domain example and a boundary-condition test case, since the primary audience is domain experts who need precise usage contracts, not implementation details. \textit{Documentation deficiency} ($31.8\%$) overtakes \textit{unimplemented feature gap} as the leading cause, an inversion indicating that documentation gaps block adoption before feature gaps do.

\emph{Ship a machine-readable harness specification document as a first-class software artifact.} RQ1 shows that $77.2\%$ of harnesses lack uncertainty quantification and $91.2\%$ lack regression alerting, yet none declare these omissions in any structured form that downstream users can inspect before selecting a harness. Following the model cards~\cite{mitchell2019model} and datasheets~\cite{gebru2021datasheets} precedent, developers could ship a machine-readable document declaring at least: \textit{stage coverage} (which of the five workflow stages the harness implements), \textit{key dependencies} (external datasets, packages, and APIs with pinned version constraints), and \textit{known challenge patterns} (root causes and stages where issues have historically concentrated). Declaring these upfront shifts breakage discovery from production to selection time.

\subsection{Implications for Harness Users}

\emph{Do not treat harness output as ground truth without independent verification.} Unlike software libraries where API contracts are clearly documented, evaluation harnesses embed task-specific scoring assumptions that are neither visible nor validated at configuration time. RQ3 shows that $43.3\%$ of \textit{algorithmic error} issues occur at individual scoring, and many arise from configuration-sensitive assumptions rather than universally broken logic. For instance, the BBQ task in LM Eval\footnotemark[14] hardcodes unknown-answer indices at positions \texttt{[2:13]} in \texttt{doc\_to\_targets}, silently misclassifying over 8{,}000 answers when the task is run on a dataset with a different answer distribution\footnote{\url{https://github.com/EleutherAI/LM Eval/issues/3226}}. The harness is internally consistent, but its assumptions do not hold for this dataset layout. RQ1 further shows that only $22.8\%$ of harnesses support uncertainty quantification, so most outputs provide no indication of whether a score difference reflects a real capability gap or a harness-specific artifact. Harness users should therefore treat any deviation from a harness's designed benchmark configuration as a configuration boundary condition and validate outputs against an independent reference before drawing conclusions, applying the same discipline as configuration testing in traditional software, where assumptions that hold within the designed envelope may break outside it.

\subsection{Implications for Researchers}

\emph{Treat evaluation engineering as a SE research problem with two concrete open directions.} Our root cause profile is dominated by specification-stage capability gaps and cross-cutting validation challenges, differing from the deployment-focused issues typical of MLOps research~\cite{paleyes2022challenges,kreuzberger2023machine}. First, \textit{validation gap} crosses all five stages (above $10\%$ each), yet existing contract-based frameworks~\cite{meyer1992applying} and data validation systems such as TFX~\cite{breck2019data} check only schema-level properties such as column types and value ranges. Evaluation harnesses require a stricter form of compatibility: a scorer expecting per-class probability distributions may silently receive argmax outputs that are type-valid but semantically mismatched, and no schema check catches this. Extending existing validation frameworks to encode and enforce task-specific semantic contracts at stage boundaries is an open research problem.

\emph{Investigate why established SE techniques require structural adaptation before they can apply to evaluation harness contexts.} RQ1 shows that uncertainty quantification ($22.8\%$ of harnesses), regression alerting ($8.8\%$), and production traffic evaluation ($7.0\%$) remain absent from more than $90\%$ of harnesses surveyed. In traditional software, low test coverage correlates with higher defect density and is measurable post-hoc from issue trackers. However, the absence of these capabilities means a class of defects never enters the issue tracker at all for harnesses. Each absent capability has a corresponding SE technique, but each technique embeds an assumption that evaluation harnesses violate: variance is internal to the test environment, passing thresholds are stable, and inputs are representative of production. For uncertainty quantification, software testing addresses stochastic outcomes through statistical hypothesis testing~\cite{arcuri2014hitchhiker}, but score variance in evaluation harnesses is driven by prompt sensitivity, sampling temperature, and dataset ordering, factors external to the test execution environment that existing statistical frameworks do not model~\cite{zhuo2024prosa}. For regression alerting, regression test selection~\cite{parry2021survey} assumes a stable passing threshold, but evaluation harness baselines shift with model, prompt, and dataset updates, leaving the regression criterion under model drift undefined. Protocols that track expected score distributions rather than fixed thresholds are needed, and no existing framework provides them. For production traffic evaluation, SE offers A/B testing and observability tooling, but both assume inputs are drawn from a live user distribution. Evaluation harness benchmarks use curated, static inputs. These techniques therefore require adaptation to treat the distribution shift between benchmark and production traffic as an explicit evaluation criterion rather than an external concern. Each gap opens a concrete SE research question: how to account for prompt sensitivity and sampling temperature in statistical significance tests, how to define a passing regression threshold when model and dataset baselines shift, and how to incorporate input distribution shift into benchmark coverage criteria.

\section{Threats to Validity}
\label{sec:threats}

\paragraph{Conclusion Validity} Our quantitative distributions reflect issue counts rather than weighted severity, potentially overrepresenting minor operational challenges relative to critical defects. Our count-based findings therefore characterize the \emph{frequency} of operational challenges but not their \emph{severity}, and the implications in \S\ref{sec:implications} should be read as identifying areas of frequent friction rather than a strict priority ordering by impact. Additionally, our statistical analyses assume independence between issues, which may not hold when multiple issues stem from the same underlying infrastructure problem. We mitigate classification reliability concerns through multi-step validation: LLM-based workflow classification achieves Cohen's $\kappa = 0.931$ on a statistically representative sample, and root cause classification achieves $\kappa = 0.873$ against human consensus labels. The lower human inter-rater agreement for root cause annotation ($\kappa = 0.758$) reflects the inherent ambiguity of attributing a single primary cause to issues that may involve interacting factors. The consensus labels used for LLM calibration resolve these disagreements through joint review: we discuss each disagreement case, present our reasoning, and reach a shared label through deliberation.

\paragraph{Construct Validity} Our classification assigns each issue to a single primary workflow stage and root cause, though operational challenges occasionally affect multiple stages or involve interacting factors. This single-label assignment simplifies complex scenarios where challenges propagate across stage boundaries. Following the annotation protocol in \S\ref{sec:method:classification_methodology}, we apply disambiguation rules to select the primary stage (earliest blocker) and root cause (most direct technical cause). During manual annotation, $14.3\%$ of sampled issues ($54$ of $377$ workflow-relevant issues) involved annotator disagreement on stage or root cause assignment. This enables clear statistical aggregation at the cost of potentially underrepresenting cascading and multi-cause challenges. In particular, the ``concentration'' of root causes in specific stages (RQ3) may be partly amplified by single-label assignment, since an issue involving both a validation gap in Specification and an algorithmic error in Assessment would be assigned to only one stage.

\paragraph{External Validity} Our study focuses on open-source harnesses hosted on GitHub. We include repositories with at least $500$ stars and active maintenance within the last 12 months, consistent with the inclusion criteria of the community-curated Awesome Production ML List used to seed and bound our dataset (\S\ref{sec:method:harness_collection}). This selection favors well-maintained, community-endorsed repositories and may exclude smaller but operationally significant harnesses used in industry. The $500$-star threshold also introduces a popularity bias: popular projects attract larger user bases that file more issues, so aggregate distributions may be disproportionately influenced by a few high-traffic repositories. To partially mitigate this concern, the archetype-level analysis in RQ2 reports \emph{normalized} issue counts (total issues divided by the number of harnesses per archetype), enabling comparison across archetypes of different sizes. We do not normalize by repository size, age, or number of contributors at the individual-harness level because our goal is to characterize what problems exist across the evaluation harness landscape (aggregate patterns), not to compare per-harness defect rates. Normalizing per harness would answer a different research question. The temporal snapshot captures a rapidly evolving ecosystem. Our findings characterize current patterns but may not generalize as evaluation infrastructure matures. However, our $57$ harnesses span diverse ML domains (language models, computer vision, reinforcement learning, speech processing), which improves broad coverage of contemporary OSS practices.

\paragraph{Internal Validity} Our issue distribution analysis may be affected by survivorship bias: users who fail at earlier stages (\eg, Provisioning) never reach later stages (\eg, Assessment), so absolute issue counts across stages reflect different user populations rather than a single cohort. Cross-stage \emph{volume} comparisons (\eg, ``Specification has $41.4\%$ of issues'') therefore reflect the \emph{observed} distribution of reported challenges, not a controlled comparison of stage difficulty. The \emph{within-stage} root cause compositions in RQ3 are less affected, as they describe the relative mix among users who \emph{do} reach each stage. Similarly, issues reported on GitHub represent only problems users chose to document publicly, excluding challenges resolved through private channels or abandoned attempts. Issue-filing culture may also vary across user communities, which could partly explain the higher normalized issue counts for Full-Stack LLM Evaluation Platforms (RQ2). The temporal aggregation of issues across harness evolution may conflate historical problems with current state. A time-windowed analysis could isolate current-state patterns but would reduce sample sizes for newer harnesses, and we leave this refinement to future work.

\section{Conclusion}
\label{sec:conclusion}

In this work, we present an empirical study of evaluation harnesses as software products. Our study establishes a workflow model comprising five stages, nine steps, and $34$ strategies; maps where developer-reported challenges concentrate across stages using $16{,}560$ classified GitHub issues; derives a root cause taxonomy of ten challenge categories spanning both defects and capability gaps; and identifies adoption gaps in production-oriented capabilities such as uncertainty quantification and regression alerting. Together, these findings establish an empirical foundation for treating evaluation reliability as a first-class concern of EvalEng. Our results indicate that improving evaluation reliability requires attention to workflow design across all stages rather than isolated metric or benchmark improvements. Two concrete directions follow from this foundation: developing structured transparency documents that report workflow coverage, key dependencies, and dominant challenge patterns for cross-harness comparison; and designing validity-oriented evaluation methods that incorporate uncertainty estimation, regression detection, and production-traffic assessment. The workflow model and root cause taxonomy we provide offer a stable baseline that longitudinal research can use to track whether adoption gaps close over time and whether adoption of capabilities such as uncertainty quantification, regression alerting, and production traffic evaluation correlates with reduced incidence in the root cause categories we report.

\bibliographystyle{ACM-Reference-Format}
\bibliography{references}

\appendix

\begin{figure}[H]
\centering
\includegraphics[width=\columnwidth]{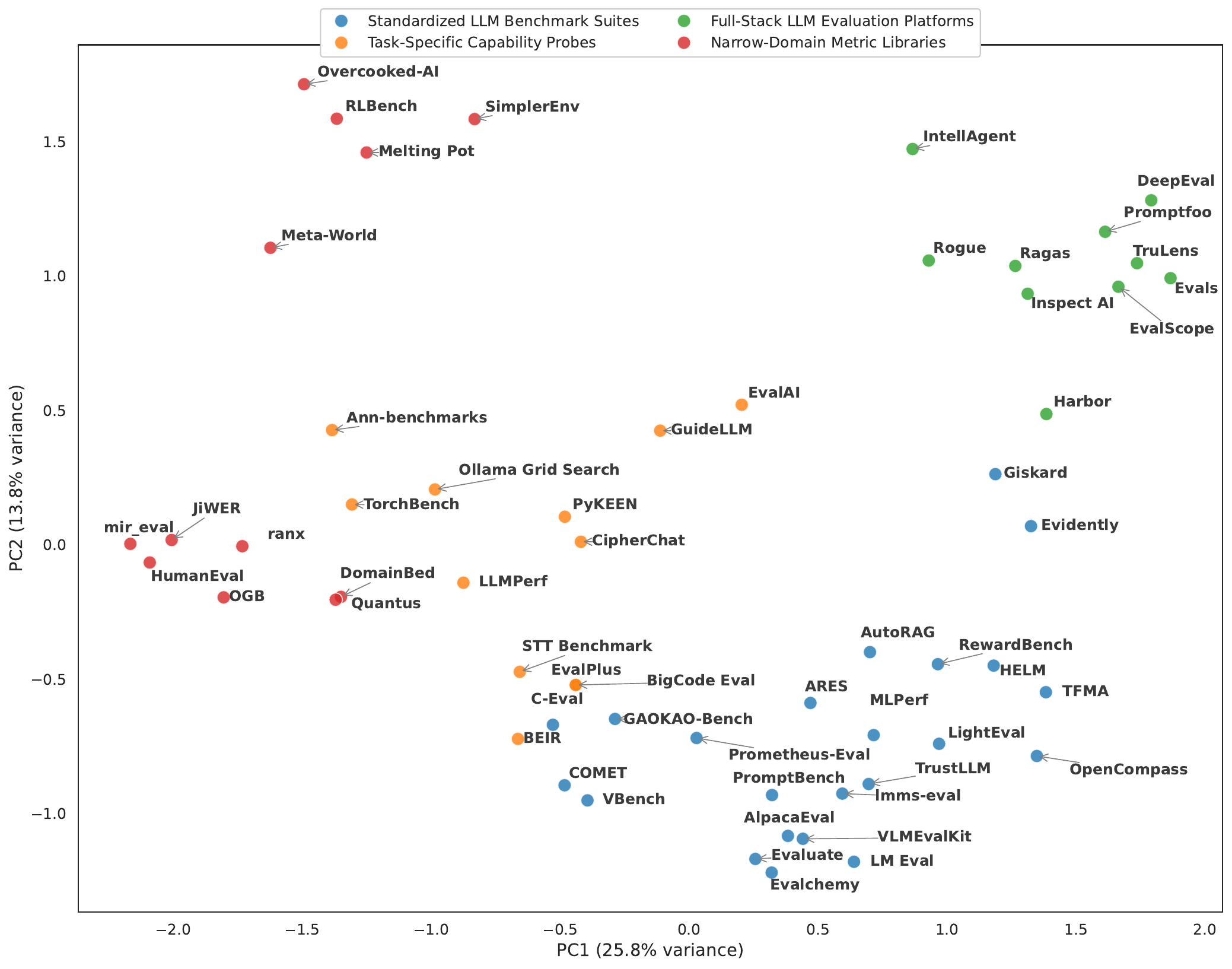}
\caption{PCA projection of $57$ harnesses based on strategy support matrix, colored by cluster membership. The distinct region occupied by each archetype reflects its operational profile, with the first principal component (x-axis) largely separating standardized benchmark suites from full-stack platforms, and the second principal component (y-axis) distinguishing narrow-domain metric libraries from task-specific probes.}
\label{fig:rq1_pca}
\end{figure}

\begin{table*}[!t]
\centering
\small
\caption{Search keywords related to evaluation tooling and their corresponding retrieved evaluation harnesses meeting inclusion criteria (500+ stars, active within 12 months).}
\label{tab:harness_search_keywords}
\begin{tabularx}{\textwidth}{llX}
Keyword & Count & Retrieved harnesses \\
\midrule
benchmark environment & 7 & Overcooked-AI\footnotemark[1], RLBench\footnotemark[2], Meta-World\footnotemark[3] \\
\midrule
benchmark framework & 44 & EvalScope\footnotemark[4], PromptBench\footnotemark[5], Speech-to-Text Benchmark\footnotemark[6], Evals\footnotemark[7] \\
\midrule
benchmark library & 31 & ANN-Benchmarks\footnotemark[8], LLMPerf\footnotemark[9] \\
\midrule
benchmark platform & 11 & OpenCompass\footnotemark[10] \\
\midrule
benchmark suite & 12 &  \\
\midrule
benchmark tool & 42 &  \\
\midrule
comparison library & 8 & ranx\footnotemark[11] \\
\midrule
comparison platform & 8 &  \\
\midrule
evals & 37 & Promptfoo\footnotemark[12], Evals\footnotemark[7], TruLens\footnotemark[13], EvalScope\footnotemark[4] \\
\midrule
evaluation environment & 10 &  \\
\midrule
evaluation framework & 39 & LM Eval\footnotemark[14], EvalScope\footnotemark[4], HELM\footnotemark[15], AutoRAG\footnotemark[16], DeepEval\footnotemark[17], Evidently\footnotemark[18], PromptBench\footnotemark[5], Evals\footnotemark[7], IntellAgent\footnotemark[19], GAOKAO-Bench\footnotemark[20], CipherChat\footnotemark[21], COMET\footnotemark[22], BigCode Eval\footnotemark[23], Inspect AI\footnotemark[24], Harbor\footnotemark[54] \\
\midrule
evaluation function & 5 & mir\_eval\footnotemark[25] \\
\midrule
evaluation harness & 2 & LM Eval\footnotemark[14], BigCode Eval\footnotemark[23] \\
\midrule
evaluation library & 35 & Evaluate\footnotemark[26], PyKEEN\footnotemark[27] \\
\midrule
evaluation notebook & 1 &  \\
\midrule
evaluation package & 7 &  \\
\midrule
evaluation platform & 18 & OpenCompass\footnotemark[10], Ollama Grid Search\footnotemark[28], GuideLLM\footnotemark[29] \\
\midrule
evaluation repository & 15 &  \\
\midrule
evaluation suite & 2 & C-Eval\footnotemark[30] \\
\midrule
evaluation tool & 28 & RewardBench\footnotemark[31] \\
\midrule
evaluation toolkit & 15 & VLMEvalKit\footnotemark[32], LightEval\footnotemark[33], TrustLLM\footnotemark[34], Quantus\footnotemark[35], Evalchemy\footnotemark[36] \\
\midrule
evaluator & 279 & OpenCompass\footnotemark[10], LM Eval\footnotemark[14], EvalScope\footnotemark[4], TruLens\footnotemark[13], DeepEval\footnotemark[17], TorchBench\footnotemark[37], Giskard\footnotemark[38], lmms-eval\footnotemark[39], VBench\footnotemark[40], VLMEvalKit\footnotemark[32], RewardBench\footnotemark[31], Evidently\footnotemark[18], LightEval\footnotemark[33], EvalAI\footnotemark[41], PromptBench\footnotemark[5], Quantus\footnotemark[35], Ollama Grid Search\footnotemark[28], Evaluate\footnotemark[26], Prometheus-Eval\footnotemark[42], GAOKAO-Bench\footnotemark[20], CipherChat\footnotemark[21], COMET\footnotemark[22], Evals\footnotemark[7], AlpacaEval\footnotemark[43], ARES\footnotemark[44], BigCode Eval\footnotemark[23], BEIR\footnotemark[45], SimplerEnv\footnotemark[46], JiWER\footnotemark[47], HumanEval\footnotemark[48], EvalPlus\footnotemark[49], OGB\footnotemark[50], HELM\footnotemark[15], AutoRAG\footnotemark[16], Inspect AI\footnotemark[24], Rogue\footnotemark[51] \\
\midrule
test framework & 205 & Evidently\footnotemark[18], Rogue\footnotemark[51] \\
\midrule
test suite & 45 & Melting Pot\footnotemark[52], DomainBed\footnotemark[53] \\
\midrule
testing tool & 254 &  \\
\end{tabularx}
\footnotetext[1]{\url{https://github.com/HumanCompatibleAI/overcooked_ai}}
\footnotetext[2]{\url{https://github.com/stepjam/RLBench}}
\footnotetext[3]{\url{https://github.com/Farama-Foundation/Metaworld}}
\footnotetext[4]{\url{https://github.com/modelscope/evalscope}}
\footnotetext[5]{\url{https://github.com/microsoft/promptbench}}
\footnotetext[6]{\url{https://github.com/Picovoice/speech-to-text-benchmark}}
\footnotetext[7]{\url{https://github.com/openai/evals}}
\footnotetext[8]{\url{https://github.com/erikbern/ann-benchmarks}}
\footnotetext[9]{\url{https://github.com/ray-project/llmperf}}
\footnotetext[10]{\url{https://github.com/open-compass/opencompass}}
\footnotetext[11]{\url{https://github.com/AmenRa/ranx}}
\footnotetext[12]{\url{https://github.com/promptfoo/promptfoo}}
\footnotetext[13]{\url{https://github.com/truera/trulens}}
\footnotetext[14]{\url{https://github.com/EleutherAI/lm-evaluation-harness}}
\footnotetext[15]{\url{https://github.com/stanford-crfm/helm}}
\footnotetext[16]{\url{https://github.com/Marker-Inc-Korea/AutoRAG}}
\footnotetext[17]{\url{https://github.com/confident-ai/deepeval}}
\footnotetext[18]{\url{https://github.com/evidentlyai/evidently}}
\footnotetext[19]{\url{https://github.com/plurai-ai/intellagent}}
\footnotetext[20]{\url{https://github.com/OpenLMLab/GAOKAO-Bench}}
\footnotetext[21]{\url{https://github.com/RobustNLP/CipherChat}}
\footnotetext[22]{\url{https://github.com/Unbabel/COMET}}
\footnotetext[23]{\url{https://github.com/bigcode-project/BigCode Eval}}
\footnotetext[24]{\url{https://github.com/UKGovernmentBEIS/inspect_ai}}
\footnotetext[25]{\url{https://github.com/mir-evaluation/mir_eval}}
\footnotetext[26]{\url{https://github.com/huggingface/evaluate}}
\footnotetext[27]{\url{https://github.com/pykeen/pykeen}}
\footnotetext[28]{\url{https://github.com/dezoito/ollama-grid-search}}
\footnotetext[29]{\url{https://github.com/vllm-project/guidellm}}
\footnotetext[30]{\url{https://github.com/hkust-nlp/ceval}}
\footnotetext[31]{\url{https://github.com/allenai/reward-bench}}
\footnotetext[32]{\url{https://github.com/open-compass/VLMEvalKit}}
\footnotetext[33]{\url{https://github.com/huggingface/lighteval}}
\footnotetext[34]{\url{https://github.com/HowieHwong/TrustLLM}}
\footnotetext[35]{\url{https://github.com/understandable-machine-intelligence-lab/Quantus}}
\footnotetext[36]{\url{https://github.com/mlfoundations/evalchemy}}
\footnotetext[37]{\url{https://github.com/pytorch/benchmark}}
\footnotetext[38]{\url{https://github.com/Giskard-AI/giskard-oss}}
\footnotetext[39]{\url{https://github.com/EvolvingLMMs-Lab/lmms-eval}}
\footnotetext[40]{\url{https://github.com/Vchitect/VBench}}
\footnotetext[41]{\url{https://github.com/Cloud-CV/EvalAI}}
\footnotetext[42]{\url{https://github.com/prometheus-eval/prometheus-eval}}
\footnotetext[43]{\url{https://github.com/tatsu-lab/alpaca_eval}}
\footnotetext[44]{\url{https://github.com/stanford-futuredata/ARES}}
\footnotetext[45]{\url{https://github.com/beir-cellar/beir}}
\footnotetext[46]{\url{https://github.com/simpler-env/SimplerEnv}}
\footnotetext[47]{\url{https://github.com/jitsi/jiwer}}
\footnotetext[48]{\url{https://github.com/openai/human-eval}}
\footnotetext[49]{\url{https://github.com/evalplus/evalplus}}
\footnotetext[50]{\url{https://github.com/snap-stanford/ogb}}
\footnotetext[51]{\url{https://github.com/qualifire-dev/rogue}}
\footnotetext[52]{\url{https://github.com/google-deepmind/meltingpot}}
\footnotetext[53]{\url{https://github.com/facebookresearch/DomainBed}}
\footnotetext[54]{\url{https://github.com/harbor-framework/harbor}}
\end{table*}

\clearpage

\setlength{\LTcapwidth}{\textwidth}
\setlength{\LTleft}{0pt}
\setlength{\LTright}{0pt}
\setlength{\tabcolsep}{3pt}

\begin{longtable}{
  >{\centering\arraybackslash}m{0.07\textwidth}|
  >{\centering\arraybackslash}m{0.17\textwidth}|
  >{\centering\arraybackslash}m{0.06\textwidth}|
  >{\centering\arraybackslash}m{0.51\textwidth}|
  >{\centering\arraybackslash}m{0.10\textwidth}
}
\caption{Workflow Component Definitions}\label{tab:workflow-components} \\
\hline
\textbf{Index} & \textbf{Name} & \textbf{\%} & \textbf{Definition} & \textbf{Example} \\
\hline
\endfirsthead
\multicolumn{5}{c}{{\tablename\ \thetable{} -- continued from previous page}} \\
\hline
\textbf{Index} & \textbf{Name} & \textbf{\%} & \textbf{Definition} & \textbf{Example} \\
\hline
\endhead
\hline \multicolumn{5}{r}{{Continued on next page}} \\
\endfoot
\hline
\endlastfoot
\multicolumn{5}{p{\dimexpr\textwidth-2\tabcolsep}}{\textbf{Stage 0: Provisioning (The Runtime)}: \emph{Establishing the technical foundation by installing required software and configuring credentials for external access.}} \\ \hline
\multicolumn{5}{p{\dimexpr\textwidth-2\tabcolsep}}{\textbf{Step S0-A: Harness Installation}: \emph{Installing dependencies, compiling binaries, building containers, and configuring execution backends.}} \\ \hline
S0-A1 & Git Clone & 100\% & Cloning the repository via git clone and installing from the cloned source. & All \\ \hline
S0-A2 & Python Package & 94.7\% & Installing Python packages via package managers including pip, uv, conda, or poetry. & OpenAI Evals \\ \hline
S0-A3 & Container Image & 21.1\% & Pulling prebuilt Docker or OCI container images that include the harness and all runtime dependencies in an isolated environment. & Promptfoo \\ \hline
S0-A4 & Binary Package & 3.5\% & Downloading standalone executable binaries that run without requiring separate dependency installation. & Ollama Grid Search \\ \hline
S0-A5 & Node Package & 1.8\% & Installing JavaScript-based harnesses via Node.js package managers including npm, npx, or system package managers, such as Homebrew. & Promptfoo \\ \hline
\multicolumn{5}{p{\dimexpr\textwidth-2\tabcolsep}}{\textbf{Step S0-B: Credential Configuration}: \emph{Authenticating with model repositories, dataset platforms, evaluation services, and leaderboard APIs.}} \\ \hline
S0-B1 & Repository Authentication & 75.4\% & Authenticating with artifact repository platforms (Hugging Face Hub, Zenodo, ModelScope), either directly or via dependency libraries, using CLI login, access tokens, or environment variables to retrieve gated/private models, datasets, and other artifacts \emph{(to S1-A1, S1-B1)}. & DeepEval \\ \hline
S0-B2 & Model API Authentication & 68.4\% & Configuring environment variables or credential files with API keys to enable remote inference requests to commercial model providers' hosted endpoints (OpenAI API, Anthropic API, HuggingFace Inference API, Google Gemini API) \emph{(to S1-A2)}. & Ragas \\ \hline
S0-B3 & Evaluation Platform Authentication & 19.3\% & Authenticating with evaluation platforms using account registration or command-line login flows to access platform services and features (configuring evaluations, running experiments, viewing results, submitting to leaderboards) \emph{(to multiple stages)}. & Giskard \\ \hline
\multicolumn{5}{p{\dimexpr\textwidth-2\tabcolsep}}{\textbf{Stage 1: Specification (The Contract)}: \emph{Defining the evaluation experiment: what to test, what to test it with, and how to judge the results.}} \\ \hline
\multicolumn{5}{p{\dimexpr\textwidth-2\tabcolsep}}{\textbf{Step S1-A: System Under Test (SUT) Preparation}: \emph{Specifying how to interact with the System Under Test (SUT), the primary algorithm, model, or system being evaluated, not auxiliary components used to test them.}} \\ \hline
S1-A1 & Model-in-Process (Local Inference) & 77.2\% & Evaluating parametric models with learned weights running on local or user-controlled infrastructure via single-shot inference where model weights are loaded into memory, enabling access to model internals (activations, logits, hidden states) \emph{(from S0-B1; to S2-A)}. & LM Eval \\ \hline
S1-A2 & Model-as-a-Service (Remote Inference) & 70.2\% & Evaluating parametric models with learned weights running on external, remotely-hosted infrastructure via single-shot HTTP endpoints, SDK clients, or API wrappers \emph{(from S0-B2; to S2-A)}. & Ragas \\ \hline
S1-A3 & Interactive Agent (Sequential Decision-Making) & 28.1\% & Evaluating stateful entities that make sequential decisions over multiple timesteps, running on local infrastructure through iterative environment observation and action selection, including reinforcement learning policies, multi-agent systems, robot controllers, and tool-using LLM agents \emph{(to S2-A2)}. & OpenAI Evals \\ \hline
S1-A4 & Non-Parametric Algorithm (Deterministic Computation) & 14.0\% & Evaluating algorithmic procedures without learned weights running on local infrastructure via single-shot computation, where deterministic algorithms operate purely on data structures and rules, including ANN algorithms (vector indexes, such as FAISS, HNSW) and ranking/retrieval algorithms (BM25, TF-IDF) \emph{(to S2-A)}. & TruLens \\ \hline
\multicolumn{5}{p{\dimexpr\textwidth-2\tabcolsep}}{\textbf{Step S1-B: Benchmark Inputs Preparation}: \emph{Acquiring and configuring the test inputs that will be used to evaluate the SUT.}} \\ \hline
S1-B1 & Benchmark Data Preparation (Offline) & 91.2\% & Preparing a predefined set of test inputs before execution, either by loading and transforming pre-existing benchmark input datasets from remote or local sources or by accepting manually specified custom test inputs, with optional preprocessing steps (data splitting, normalization, formatting) \emph{(from S0-B1; to S2-A)}. & LM Eval \\ \hline
S1-B2 & Synthetic Data Generation (Generative) & 40.4\% & Creating test data on the fly through input perturbation, test augmentation, trajectory generation, and scenario synthesis \emph{(to S2-A)}. & DeepEval \\ \hline
S1-B3 & Simulation Environment Setup (Simulated) & 14.0\% & Initializing interactive environment state through scene construction (instantiating 3D virtual environments, configuring object layouts and initial conditions, selecting goal configurations from task distributions, and assigning cooperative or adversarial agents) \emph{(to S2-A2)}. & Metaworld \\ \hline
S1-B4 & Production Traffic Sampling (Online) & 7.0\% & Sampling real-world inference traffic for evaluation through stream buffering and feedback collection \emph{(to S2-A4)}. & Evidently \\ \hline
\multicolumn{5}{p{\dimexpr\textwidth-2\tabcolsep}}{\textbf{Step S1-C: Benchmark References Preparation}: \emph{Pre-computing judges, references, and ground truth materials that will be used to score SUT invocation outputs.}} \\ \hline
S1-C1 & Ground Truth Preparation & 91.2\% & Pre-loading and pre-computing ground truth reference materials including human annotations, embedding indexes, extracted knowledge claims, model attribution saliency maps, statistical baseline features, and ranking ground truths \emph{(to S3-A, S3-B1)}. & LM Eval \\ \hline
S1-C2 & Judge Preparation & 61.4\% & Setting up evaluation judge models by training specialized judges through fine-tuning discriminative or reward models on labeled preference data, quality ratings, or correctness annotations, or by loading and configuring pre-trained judge models for evaluation tasks \emph{(to S3-A2)}. & AutoRAG \\ \hline
\multicolumn{5}{p{\dimexpr\textwidth-2\tabcolsep}}{\textbf{Stage 2: Execution (The Run)}: \emph{Observing SUT behavior by applying test inputs to elicit outputs and actions.}} \\ \hline
\multicolumn{5}{p{\dimexpr\textwidth-2\tabcolsep}}{\textbf{Step S2-A: SUT Invocation}: \emph{Running the System Under Test to generate outputs or take actions.}} \\ \hline
S2-A1 & Batch Inference & 94.7\% & Execute multiple input samples through a single SUT instance via configurable invocation strategies ranging from direct model calls to sophisticated multi-step architectures (prompt engineering, retrieval augmentation, multi-turn dialog, agent scaffolds), running separate evaluation runs for each SUT when evaluating multiple systems \emph{(from S1-A, S1-B1/B2; to S3-A)}. & OpenAI Evals \\ \hline
S2-A2 & Interactive Loop & 31.6\% & Statefully stepping through state transitions via iterative SUT actions through tool-based reasoning, physics simulation, and multi-agent coordination \emph{(from S1-A3, S1-B3; to S3-A, S4-A5)}. & Metaworld \\ \hline
S2-A3 & Arena Battle & 12.3\% & Execute the same input sample across multiple SUTs simultaneously in a single execution run, producing paired outputs for direct comparison \emph{(from S1-A; to S3-A2)}. & DeepEval \\ \hline
S2-A4 & Production Streaming & 7.0\% & Continuously processing live production traffic with real-time metric collection via drift monitoring and interactive feedback \emph{(from S1-B4; to S3-A, S4-A6)}. & Evidently \\ \hline
\multicolumn{5}{p{\dimexpr\textwidth-2\tabcolsep}}{\textbf{Stage 3: Assessment (The Score)}: \emph{Converting observations into measurements: judging outputs against quality criteria to produce scores.}} \\ \hline
\multicolumn{5}{p{\dimexpr\textwidth-2\tabcolsep}}{\textbf{Step S3-A: Individual Scoring}: \emph{Computing metrics for individual test instances based on SUT outputs.}} \\ \hline
S3-A1 & Deterministic Measurement & 89.5\% & Direct rule-based calculations performed without embedding transformation, including equality checks (unit test pass/fail, answer extraction), distance metrics (edit distance, geometric distance), and token-based text metrics (BLEU, ROUGE, METEOR) \emph{(from S2-A, S1-C1; to S3-B, S4-A)}. & LM Eval \\ \hline
S3-A2 & Subjective Measurement & 59.6\% & Model-based judgments with inherent uncertainty, using LLMs or classifiers as evaluators to assess subjective attributes that would typically require human judgment, including pairwise comparison of outputs from different SUTs \emph{(from S2-A, S1-C2; to S3-B, S4-A)}. & LM Eval \\ \hline
S3-A3 & Latent Measurement & 49.1\% & Semantic similarity and alignment calculations requiring transformation into a learned latent space (embedding space) where semantically similar items are positioned closer together within a continuous manifold, enabling distance-based comparisons (cosine similarity, BERTScore) \emph{(from S2-A, S1-C1; to S3-B, S4-A)}. & LM Eval \\ \hline
S3-A4 & Performance Measurement & 38.6\% & Measuring resource consumption and efficiency tradeoffs, including time costs (latency, throughput), computational costs (memory, FLOPs), and energy costs (power consumption, carbon footprint) \emph{(from S2-A; to S3-B, S4-A)}. & Promptfoo \\ \hline
\multicolumn{5}{p{\dimexpr\textwidth-2\tabcolsep}}{\textbf{Step S3-B: Aggregate Scoring}: \emph{Aggregating instance-level scores into benchmark-level metrics, a fundamental operation supported by all evaluation harnesses.}} \\ \hline
S3-B1 & Distributional Statistics & 96.5\% & Computing benchmark-level metrics from per-instance scores using averaging and quantiles, weighted aggregation, metric fusion, and rank aggregation \emph{(from S3-A, S1-C1; to S4-A)}. & OpenAI Evals \\ \hline
S3-B2 & Uncertainty Quantification & 22.8\% & Estimating confidence bounds around aggregate metrics using bootstrap resampling or Prediction-Powered Inference (PPI) that combines labeled and unlabeled data \emph{(from S3-A; to S4-A)}. & LM Eval \\ \hline
\multicolumn{5}{p{\dimexpr\textwidth-2\tabcolsep}}{\textbf{Stage 4: Reporting (The Output)}: \emph{Making results actionable: translating metrics into stakeholder-facing insights.}} \\ \hline
\multicolumn{5}{p{\dimexpr\textwidth-2\tabcolsep}}{\textbf{Step S4-A: Insight Presentation}: \emph{Visualizing metrics and publishing results to internal/external audiences.}} \\ \hline
S4-A1 & Chart Generation & 43.9\% & Creating visual representations including radar charts for multi-dimensional quality profiles, drift histograms showing distribution changes, and performance trend plots \emph{(from S3-B)}. & DeepEval \\ \hline
S4-A2 & Dashboard Creation & 45.6\% & Building interactive web interfaces displaying metric comparisons, ranked result tables, and filterable evaluation outcomes \emph{(from S3-B)}. & LM Eval \\ \hline
S4-A3 & Leaderboard Publication & 40.4\% & Submitting evaluation results to public or private leaderboards for SUT comparison \emph{(from S3-B)}. & LM Eval \\ \hline
S4-A4 & Subgroup Analysis & 40.4\% & Breaking down aggregate performance metrics by demographic groups, data domains, task categories, or other stratification dimensions \emph{(from S3-A)}. & DeepEval \\ \hline
S4-A5 & Execution Tracing & 33.3\% & Capturing and displaying detailed step-by-step execution logs showing intermediate computational states, function calls, data transformations, and execution flow of the SUT during test runs, with configurable recording mechanisms for persisting trajectory data \emph{(from S2-A)}. & DeepEval \\ \hline
S4-A6 & Regression Alerting & 8.8\% & Automatically comparing current evaluation results against historical baselines to detect performance degradation and trigger alerts when metrics fall below defined thresholds \emph{(from S2-A4, S3-B)}. & TorchBench \\
\end{longtable}

\end{document}